\documentstyle[11pt,aaspp4,psfig]{article}

\def\gs{\mathrel{\raise0.27ex\hbox{$>$}\kern-0.70em 
\lower0.71ex\hbox{{$\scriptstyle \sim$}}}}
\def\ls{\mathrel{\raise0.27ex\hbox{$<$}\kern-0.70em 
\lower0.71ex\hbox{{$\scriptstyle \sim$}}}}

\tighten

\begin{document}
\renewcommand{\thefootnote}{\fnsymbol{footnote}}
\title{
Measurements of the Cosmological Parameters $\Omega$ and $\Lambda$\\
from the First 7 Supernovae at $z \ge 0.35$\footnote{Based in part on data
from the Isaac Newton Group Telescopes, KPNO and CTIO Observatories run by
AURA, Mount Stromlo \& Siding Spring Observatory, Nordic Optical Telescope, and
the W. M. Keck Observatory}\\
}
\renewcommand{\thefootnote}{\arabic{footnote}}


\author{ 
S.~Perlmutter,\altaffilmark{1,2}
S.~Gabi,\altaffilmark{1,3}
G.~Goldhaber,\altaffilmark{1,2}
A.~Goobar,\altaffilmark{1,2,4}
D.~E.~Groom,\altaffilmark{1,2}
I.~M.~Hook,\altaffilmark{2,10}
A.~G.~Kim,\altaffilmark{1,2} 
M.~Y.~Kim,\altaffilmark{1}
J.~C.~Lee,\altaffilmark{1}
R.~Pain,\altaffilmark{1,5}
C.~R.~Pennypacker,\altaffilmark{1,3}
I.~A.~Small,\altaffilmark{1,2} 
R.~S.~Ellis,\altaffilmark{6}
R.~G.~McMahon,\altaffilmark{6} 
B.~J.~Boyle,\altaffilmark{7,8}
P.~S.~Bunclark,\altaffilmark{7}
D.~Carter,\altaffilmark{7} 
M.~J.~Irwin,\altaffilmark{7}
K.~Glazebrook,\altaffilmark{8}
H.~J.~M.~Newberg,\altaffilmark{9}
A.~V.~Filippenko,\altaffilmark{2,10}
T.~Matheson,\altaffilmark{10}
M.~Dopita,\altaffilmark{11}
and W.~J.~Couch\altaffilmark{12}\\
(The Supernova Cosmology Project)\\
\mbox{}\\
\mbox{}\\
Accepted by {\em The Astrophysical Journal}.
}
\altaffiltext{1}{Institute for Nuclear and Particle Astrophysics,
E.~O. Lawrence Berkeley National Laboratory, 50-232,
Berkeley, California 94720; saul@lbl.gov}
\altaffiltext{2}{Center for Particle Astrophysics, 
    U.C. Berkeley, California 94720}
\altaffiltext{3}{Space Sciences Laboratory, U.C. Berkeley, California 94720}
\altaffiltext{4}{University of Stockholm}
\altaffiltext{5}{LPNHE, CNRS-IN2P3 and Universit\'es Paris VI \& VII, France }
\altaffiltext{6}{Institute of Astronomy, Cambridge, United Kingdom}
\altaffiltext{7}{Royal Greenwich Observatory, Cambridge, United Kingdom}
\altaffiltext{8}{Anglo-Australian Observatory, Sydney, Australia}
\altaffiltext{9}{Fermilab, Batavia, Illinois 60510}
\altaffiltext{10}{Department of Astronomy, University of California, Berkeley, California 94720-3411}
\altaffiltext{11}{Mt. Stromlo and Siding Springs Observatory, Australia}
\altaffiltext{12}{University of New South Wales, Sydney, Australia}


\begin{abstract}
We have developed a technique to systematically discover and study
high-redshift supernovae that can be used to measure the cosmological
parameters. We report here results based on the initial seven of $>$28
supernovae discovered to date in the high-redshift supernova search of
the Supernova Cosmology Project.   We find an observational dispersion in peak
magnitudes of $\sigma_{M_B} = 0.27$; this dispersion narrows
to $\sigma_{M_B,{\rm corr}} = 0.19$ after ``correcting'' the
magnitudes using the light-curve ``width-luminosity'' relation found for
nearby ($z \le 0.1$) type Ia supernovae from the Cal\'{a}n/Tololo survey 
(Hamuy {\em et~al.} 1996).  Comparing  lightcurve-width-corrected
magnitudes as a function of redshift of our distant ($z = 0.35$--0.46)
supernovae to those of nearby type Ia supernovae yields a global
measurement of  the mass density, $\Omega_{\rm M} = 0.88
\;^{+0.69}_{-0.60}$ for a $\Lambda = 0$ cosmology.  For a
spatially flat universe  (i.e., $\Omega_{\rm M} +\Omega_\Lambda = 1$),
we find $\Omega_{\rm M} = 0.94 \;^{+0.34}_{-0.28}$ or, equivalently, a
measurement of the cosmological constant, $\Omega_\Lambda = 0.06
\;^{+0.28}_{-0.34}$  ($<$0.51 at the 95\% confidence level).
For the more general Friedmann-Lema\^{\i}tre cosmologies with
independent $\Omega_{\rm M}$ and $\Omega_\Lambda$, the results are
presented as a confidence region on the $\Omega_{\rm
M}$--$\Omega_\Lambda$ plane.  This region does not correspond to a
unique  value of the deceleration parameter $q_0$. We present analyses
and checks for statistical and systematic errors, and also show that our
results do not depend on the specifics of the width-luminosity
correction. The results for  $\Omega_\Lambda$-versus-$\Omega_{\rm M}$
are  inconsistent with $\Lambda$-dominated, low density, flat
cosmologies that have been proposed to reconcile the  ages of globular
cluster stars with higher Hubble constant values.  
\end{abstract}  

\keywords{cosmology: observations---distance scale---supernovae:general}

%
%

\section{Introduction}

The classical magnitude-redshift diagram for a distant standard candle
remains perhaps the most direct approach for measuring the cosmological
parameters that determine the fate of the cosmic expansion (Sandage
1961, 1989). The first standard candles used in such studies  were  
first-ranked cluster galaxies (Gunn \& Oke 1975, Kristian,  Sandage, \&
Westphal 1978) and the characteristic magnitude of the  cluster galaxy
luminosity function (Abell 1972).   More recent measurements have used
powerful radio galaxies at higher redshifts (Lilly \& Longair 1984, 
Rawlings {\em et~al.} 1993).  Both the early programs (reviewed by
Tammann 1983)  and the  recent work  have proven particularly important
for the understanding of galactic  evolution, but are correspondingly
more difficult to interpret as measurements of cosmological  parameters.
The type Ia supernovae (SNe Ia), the  brightest, most homogeneous class
of supernovae,  offer an attractive alternative candle,  and have
features that address this evolution problem.   Each supernova explosion
emits a  rich stream of information describing the event, which we
observe  in the form of multi-color light curves and time-varying
spectra.  Supernovae at high redshifts, unlike galaxies, are events
rather than objects, and their detailed temporal behavior can thus be
studied on an individual basis for signs of evolution relative to nearby
examples.  

The disadvantages of using supernovae are also obvious:  they are rare,
transient events that occur at unpredictable times, and are therefore
unlikely candidates for the scheduled observations necessary on the
largest telescopes.  The single previously identified high-redshift ($z
= 0.31$) SN~Ia, discovered by a 2-year Danish/ESO search in Chile, was
found (at an unpredictable time) several weeks after it had already
passed  its peak luminosity (N{\o}rgaard-Nielsen {\em et~al.} 1989).

To make high-redshift supernovae a more practical ``cosmological tool,''
the Supernova Cosmology Project has developed a technique over the past
several years  that allows the discovery of high-redshift SNe Ia in
groups of ten or more at one time (Perlmutter {\em et~al.} 1997a). 
These ``batch'' discoveries are scheduled for a particular night, or
nights, thus also allowing follow-up spectroscopy and photometry on the
large-aperture  telescopes to be scheduled.  Moreover, the supernova
discoveries are generally selected to be on the  rising part of the
light curves, and can be chosen to occur just before new moon for
optimal observing conditions at maximum light.

Since our demonstration of this technique with the discovery of SN
1992bi at $z=0.458$  (Perlmutter {\em et~al.} 1994, 1995a), we have now
discovered more than 28 SNe, most in two batches of $\sim $10
(Perlmutter {\em et~al.} 1995b, 1997b).  Almost all  are SNe Ia detected
before maximum light in the redshift range $z = 0.35$--0.65.  We have 
followed all of these supernovae with photometry and almost all with
spectroscopy, usually at the Keck 10-meter telescope.    Other groups
have now begun high-redshift searches; in particular, the search of
Schmidt {\em et~al.} (1997) has recently  reported the discovery of
high-redshift supernovae (Garnavich {\em et~al.} 1996a,b).

We report here the measurements of the cosmological parameters from the
initial seven supernovae discovered at redshifts $z \ge 0.35$.  Since
this is a first measurement using this technique, we present some detail
to define terms, demonstrate cross-checks of the measurement,  and 
outline the directions for future refinements. Section 2 reviews the
basic equations of the technique and defines useful variables that are
independent of $H_0$.   Section 3 discusses the current understanding of
type Ia supernovae as calibrated standard candles, based on low-redshift
supernova studies.  Section 4 describes the high-redshift supernova data
set.   Section 5 presents several different analysis approaches all
yielding essentially the same results.  Section 6 lists checks for
systematic errors, and shows that none of these sources of error will
significantly change the current results.  

In conclusion (Section 7), we find that the  alternative analyses and
cross-checks for systematic error all provide confidence in this
relatively simple measurement, a magnitude versus redshift, that gives
an independent measurement of $\Omega_{\rm M}$ and $\Omega_\Lambda$
comparable to or better than previous measurements and limits. Other
current and forthcoming papers discuss further scientific results from
this data set and provide catalogs of light curves and spectra: Pain
{\em et~al.} (1996) present first evidence that high-redshift SN~Ia
rates are comparable to low-redshift rates,  Kim {\em et~al.} (1996)
discuss implications for the Hubble constant,  and Goldhaber {\em
et~al.} (1996) present evidence for time dilation of events at high
redshift.

\section{Measurement of $\Omega_{\Lambda}$ versus $\Omega_{\rm M}$ from
$m$--$z$ relation}

The classical magnitude-redshift test takes advantage of  the
sensitivity of the apparent magnitude-redshift relation to the
cosmological model.  Within Friedmann-Lema\^{\i}tre cosmological models,
the apparent bolometric magnitude $m(z)$ of a standard candle (absolute
bolometric magnitude $M$) at a given redshift is a function of {\em
both} the cosmological-constant energy density  $\Omega_\Lambda \equiv
\Lambda/(3H_0^2)$  and the mass density $\Omega_{\rm M}$:
\begin{eqnarray}
m(z) & = & M + 5 \log {d}_L(z;\Omega_{\rm M},\Omega_\Lambda,H_0) + 
   25\nonumber\\
   & \equiv & M  + 5 \log {\cal D}_L(z;\Omega_{\rm M},\Omega_\Lambda) 
   - 5\log H_0 + 25\;,
\label{magzrel}
\end{eqnarray}
where $d_L$ is the luminosity distance and ${\cal D}_L \equiv H_0 d_L$ 
is the part of the luminosity distance expression\footnotemark\ that
remains  after multiplying out the dependence on the Hubble constant 
(expressed here in units of km s$^{-1}$ Mpc$^{-1}$). In the low redshift
limit, Equation~\ref{magzrel} reduces to the usual linear Hubble
relation between $m$ and $\log cz$:
\begin{eqnarray}
m(z) & = & M  + 5 \log cz - 5\log H_0 + 25\nonumber\\
    & = & {\cal M} + 5 \log cz\;,
\label{hubblerel}
\end{eqnarray}
where we have expressed the intercept of the Hubble line as the
magnitude ``zero point'' ${\cal M} \equiv M - 5\log H_0$ + 25. This
quantity can be measured from the apparent magnitude and redshift of
low-redshift examples of the standard candle, without knowing $H_0$.
Note that the dispersions of ${\cal M}$ and $M$ are the same, 
$\sigma_{\cal M} = \sigma_M$, since $ 5\log H_0$ is constant. (In this
paper, we use script letters to represent variables that are independent
of $H_0$; the measurement of $H_0$ requires extra information, the
absolute distance to one of the standard candles, that we do not need
for the measurement of the other cosmological parameters.)

\footnotetext{We reproduce the equation for the luminosity distance,
both for completeness and to  correct a typographical error
in Goobar \& Perlmutter (1995):
\begin{equation}
d_L(z;\Omega_M,\Omega_\Lambda, H_0) =
\frac{c(1 + z)}{H_0 \sqrt{|\kappa| }} \; \; \; {\cal S}\! \left (
  \sqrt{|{\kappa}| } \int_0^{z} \left [(1+z^\prime)^2(1+\Omega_M z^\prime)-
   z^\prime (2+z^\prime ) \Omega_\Lambda \right]^{-\frac{1}{2}} dz^\prime
  \right ),
\end{equation}
where, for $\Omega_M + \Omega_\Lambda > 1$, ${\cal S}(x)$ is defined as
$\sin(x)$ and $\kappa = 1 - \Omega_M - \Omega_\Lambda $; for $\Omega_M +
\Omega_\Lambda < 1$, ${\cal S}(x) = \sinh(x)$ and $\kappa$ as above; and
for $\Omega_M + \Omega_\Lambda = 1$, ${\cal S}(x) = x$ and $\kappa =1$.
The  greater-than and less-than signs were interchanged in the
definition of  ${\cal S}(x)$  in the printed version of  Goobar \&
Perlmutter, although all calculations were performed with the correct
expression. }

Thus, with a set of apparent magnitude and redshift measurements
($m(z)$) for high-redshift candles,  and a similar set of low-redshift
measurements to determine ${\cal M}$, we can find the best fit values of 
$\Omega_{\Lambda}$ and $\Omega_{\rm M}$ to solve the equation
\begin{equation}
   m(z) - {\cal M} = 5 \log {\cal D}_L(z;\Omega_{\rm M},\Omega_\Lambda)\;.
   \label{simplemagz}
\end{equation}
(An equivalent procedure would be to fit the low- and high-redshift
measurements simultaneously to Equation~\ref{simplemagz}, leaving ${\cal
M}$ free as a fitting parameter.) For candles at a given redshift, this
fit yields a confidence region that appears as a diagonal strip on the
plane of  $\Omega_\Lambda$  versus $\Omega_{\rm M}$. Goobar \&
Perlmutter (1995) emphasized that the  integrand for the luminosity 
distance ${\cal D}_L$ depends on $\Omega_{\rm M}$ and  $\Omega_\Lambda$
with different functions of  redshift,$^{13}$ so that the slope of the
confidence-region strip increases with redshift. This change in slope
with redshift makes possible a future measurement of  both $\Omega_{\rm
M}$ and  $\Omega_\Lambda$ separately, using supernovae at a range of
redshifts from $z=0.5$ to 1.0; see Figure 1 of  Goobar \& Perlmutter.

Traditionally, the magnitude-redshift relation for a standard candle has
been interpreted as a measurement of the deceleration parameter,  $q_0$,
primarily in   the special case of a $\Lambda=0$ cosmology where $q_0$
and  $\Omega_{\rm M}$ are equivalent parameterizations of the model.
However, in the most general case, $q_0$ is a poor description of the
measurement, since ${\cal D}_L$ is a function of $\Omega_{\rm M}$ and
$\Omega_\Lambda$  independently, not simply the combination $q_0 \equiv
\Omega_{\rm M}/2 - \Omega_\Lambda$. Thus the slope of the
confidence-region strip is not parallel to contours of constant $q_0$,
except at redshifts $z <\!< 1$. We therefore recommend that for
cosmologies with a non-zero cosmological constant, $q_0$ {\em not be
used by itself to describe the measurements of the cosmological
parameters from the magnitude-redshift relation at high redshifts,}
since it will lead to confusion in the literature. 

Steinhardt (1996) recently pointed out that cosmological models can be
constructed with other forms of energy density besides $\Omega_{\rm M}$
and $\Omega_\Lambda$, such as the energy density due to
topological defects.  These energy density terms will not in general
lead to the same functional dependence of luminosity distance on redshift.
In this current paper, we do not address these cosmologies with additional
(or different) energy density terms, since this first set of high-redshift
supernovae span a relatively narrow range of redshifts.  Although
constraints on these cosmological models can be found from this limited
data, our upcoming larger data sets with a larger redshift range will be
much more appropriate for this purpose.

\section{Low-Redshift SNe Ia and Calibrated Magnitudes}

To measure the magnitude ``zero point'' ${\cal M}$, it is important to
use low-redshift supernovae that are far enough into the Hubble flow
that their peculiar velocities are not an appreciable contributor to the
redshift.  It is also better if the sample of  low-redshift supernovae
were discovered in a systematic search, since this more closely
approximates the high-redshift sample; our high-redshift  search
technique yields a more uniform (and measurable) magnitude limit than
typical of most serendipitous supernova discoveries.

The Cal\'{a}n/Tololo supernova search has discovered and followed a
sample of 29 supernovae in the redshift range $z=0.01$--0.10 (Hamuy {\em
et~al.} 1995, 1996).  Of these, 18 were discovered within 5 days of
maximum light or sooner.  This subsample is the best to use for
determining ${\cal M}$, since there is little or no extrapolation in the
measurements of the peak apparent magnitude or the light curve decline
rate. The absolute $B$-magnitude distribution of these 18
Cal\'{a}n/Tololo supernovae exhibits a relatively narrow RMS dispersion, 
$\sigma_{{M}_B}^{\rm Hamuy} = 0.26$~mag, with a mean magnitude zero
point of ${\cal M}_B = -3.17 \pm 0.03$   (Hamuy {\em et~al.}
1996).  

Recent work on samples of SNe Ia at redshifts $z\le 0.1$ has focussed
attention on examples of differences within the SN~Ia class, with
observed deviations in luminosity at maximum light, color, light curve
width, and spectrum (e.g. Filippenko {\em et~al.} 1992a,b; Phillips {\em
et~al.} 1992; Leibundgut {\em et~al.} 1993; Hamuy {\em et~al.} 1994). 
It appears that these deviations are generally highly correlated,
and---perhaps surprisingly---approximately define a single-parameter
supernova family, presumably of different explosion strengths, that may
be characterized by any of these correlated observables.  It thus seems
possible to predict, and hence correct, a deviation in luminosity using
such indicators as light curve width (Phillips 1993; Hamuy {\em et~al.}
1995; Riess, Press, \& Kirshner 1995), $U$$-$$B$ color (Branch, Nugent, \& 
Fisher 1997), or spectral feature ratios (Nugent {\em  et~al.} 1996). 

\subsection{Light Curve Width Calibration}

The dispersion $\sigma_{{M}_B}^{\rm Hamuy} = 0.26$~mag thus can be
improved by ``calibrating,''  using the correlation between the time
scale of the supernova light curve and the peak luminosity of the
supernova.  The correlation is  in the sense that broader, slower light
curves are brighter while  narrower, faster light curves are fainter. 
Phillips (1993) proposed a simple linear relationship between the
decline rate $\Delta m_{15}$, the magnitude change in the first 15 days
past $B$ maximum, and the peak $B$ absolute magnitude.  (In practice,
low-redshift supernova light curves are not always observed during these
15 days, and therefore the photometry data are fit to a series of
alternative template SN~Ia light curves that span a range of decline
rates;  $\Delta m_{15}$ is actually found by interpolating between the 
$\Delta m_{15}$ values of the templates that fit best.) Hamuy {\em
et~al.} (1995, 1996) have now fitted a linear relation for all of the 18  Cal\'{a}n/Tololo
supernovae that were discovered near maximum brightness, and  for the
observed range of  $\Delta m_{15}$ between 0.8 and 1.75 mag they obtain:

\begin{equation}
   {\cal M}_{B,{\rm corr}}  = 
   (0.86 \pm 0.21) (\Delta m_{15} - 1.1) - (3.32 \pm 0.05) \;.
\label{widthbrightrel}
\end{equation}

This fit provides a prescription for ``correcting'' magnitudes to make
them comparable to an arbitrary ``standard'' SN~Ia light curve of width
$\Delta m_{15} = 1.1$~mag:  Add the correction term $\Delta_{\rm
corr}^{\scriptscriptstyle\{1.1\}} = (-0.86 \pm 0.21) (\Delta m_{15} -
1.1)$ to the measured $B$ magnitude, so that $m_{B,{\rm corr}} = m_B +
\Delta_{\rm corr}^{\scriptscriptstyle\{1.1\}}$. (We use the \{1.1\}
superscript as a reminder that this correction term is  defined for the
arbitrary choice of light curve width, $\Delta m_{15} = 1.1$~mag.) The
residual magnitude dispersion after adding this correction to the
Cal\'{a}n/Tololo supernova magnitudes drops from  $\sigma_{{M}_B}^{\rm
Hamuy} = 0.26$ to  $\sigma_{{M}_B,{\rm corr}}^{\rm Hamuy} = 0.17$ 
magnitudes.  It is important to notice that the magnitude zero point,
${\cal M}_{B} = -3.17 \pm 0.03$, calculated from the uncorrected
magnitudes   is {\em not} the same as ${\cal M}_{B,{\rm
corr}}^{\scriptscriptstyle\{1.1\}} = -3.32 \pm 0.05 $, the  intercept of
Equation~\ref{widthbrightrel} at $\Delta m_{15} = 1.1$~mag; this simply
reflects the fact that $\Delta m_{15} = 1.1$ is not the value for the
average SN~Ia. 

Riess, Press, \& Kirshner (1995) have presented a different analysis of
this light curve width-luminosity correlation that adds or subtracts
different amounts of a ``correction template'' to a standard light-curve
template (Leibundgut {\em et~al.} 1991),  creating a similar family of
broader and narrower light curves. They use a simple linear
relationship between the amount of this correction template added and
the absolute magnitude of the supernova, resulting in  a similarly small
dispersion in the $B$ and $V$ absolute magnitude after
correction, $\sigma_{M_{B,V}}^{\rm RPK1} \approx 0.20$~mag.   More
recent results of Riess, Press, \& Kirshner (1996) show even smaller
dispersion  ($\sigma_{M_{B,V}}^{\rm RPK2} \approx 0.12$~mag) if
multiple color light curves are used and extinction terms included in
the fit.

There remains some question concerning  the details of the light curve
width-luminosity relationship.  It is not clear that a straight line is
the ``true'' model relating $\Delta m_{15} $ to $M_B$, nor that a linear
addition or subtraction of a Riess {\em et~al.} correction template best
characterizes the range of light curves in all bands.  However, a simple
inspection of the absolute magnitude as a function of $\Delta m_{15}$
from Hamuy {\em et~al.} (1995, 1996) shows primarily a narrow $M_B$
dispersion ($\sigma_{M_B} \approx 0.2$~mag) for most of the
supernovae, those with light curve widths near that of the Leibundgut
standard template ($\Delta m_{15} = 1.1$~mag), as well as a few slightly
brighter, broader supernovae and a tail of fainter, narrower supernovae. 
For the purposes of this paper, a simple linear fit appears to be
sufficient, since the differences from a more elaborate fit are  well
within the photometry errors.

To make our results robust with respect to this correction, we have
analyzed the data (a) as measured, i.e., with no correction for the
width-luminosity relation, adopting the uncorrected ``zero point'' of
Hamuy {\em et~al.}, ${\cal M}_B  = -3.17 \pm 0.03$; and (b) with the
correction and zero point of Equation~\ref{widthbrightrel} for the five
supernovae that can be corrected with the Hamuy {\em et~al.} 
calibration.  We have also compared the ``correction template'' approach
for the supernova for which photometry  was obtained in bands suitable
for this method, following the prescription and template light curves of 
Riess {\em et~al.} (1996).

\subsection{Stretch-Factor Parameterization}

For the analysis of $\Delta m_{15}$ in this paper, we use a third 
parameterization of the light curve width/shape, a stretch factor $s$
that linearly broadens or narrows the rest-frame timescale of  an
average (e.g. Leibundgut {\em et~al.} 1991) template light curve.  This
stretch factor was proposed (Perlmutter 1997a) as a simple  heuristic
alternative to using a family of light curve templates, since it fits
almost all supernova light curves with a dispersion of $<0.05$
mag at any given time in the light curve during the best-measured period
from 10 days before to 80 days after maximum light.  
(Physically, the stretch factor may reflect a temperature-dependent
variation in opacities and hence the diffusion time-scale of the
supernova atmosphere; see Khohklov, M\"uller, \& H\"oflich 1993.)

The stretch factor $s$ can be translated to a corresponding $\Delta
m_{15}$ via the best-fit line
\begin{equation}
  \Delta m_{15} = (1.96\pm 0.17)(s^{-1} - 1) + 1.07 \;.
  \label{stodeltam15}
\end{equation}
Using this relation, the best-fit $s$-factor for the template supernovae
used by Hamuy {\em et~al.} reproduces their $\Delta m_{15}$ values
within $\pm$0.01 mag for the range  $0.8 \le \Delta m_{15} \le 1.75$~mag
covered by the 18 low-redshift supernovae.  This provides a simple route
to interpolating  $\Delta m_{15}$ for  supernovae that fall between
these templates.

In our analysis, we  use Equation~\ref{stodeltam15} together with the
Hamuy {\em et~al.} width-luminosity relation
(Equation~\ref{widthbrightrel})  to calculate  the magnitude correction
term, $\Delta_{\rm corr}^{\scriptscriptstyle\{1.1\}}$,  from the stretch
factor, $s$.  Note that Equation~\ref{widthbrightrel} is based on the
$\Delta m_{15}$ interpolations of Hamuy {\em et~al.}, {\em not} on a
direct measurement of $s$, and it will be recalculated once the Hamuy
{\em et~al.} light curves are published.   However, for available light
curves   we get close agreement  (within approximately 0.04 mag) between
published  $\Delta m_{15}$ values interpolated between light curve
templates by Hamuy {\em et~al.} and the values  interpolated using $s$
and  Equation~\ref{stodeltam15}.   The uncertainty introduced by this
translation is much smaller than the uncertainties in the measurement of
$s$ for the high-redshift supernovae and the uncertainties from
approximating the width-luminosity relation  as a straight line in
Equation~\ref{widthbrightrel}.

\subsection{Color and Spectroscopic Feature Calibrators}

In addition to these parameterizations of the light-curve width or
shape, several other observable features appear to be correlated with
the absolute magnitude of the supernova.  Vaughan {\em et~al.} (1995)
suggested that a color restriction $B-V < 0.25$~mag eliminated the
subluminous supernovae from a sample of low-redshift supernovae, and 
Vaughan, Branch, \& Perlmutter (1996) confirmed this  with a more recent
sample of supernovae.  Branch {\em  et~al.} (1996) presented a
potentially stronger correlation with $U-B$ color, and showed  a very
small dispersion in graphs of $M_B$ or $\Delta m_{15}$ versus $U-B$. 
This result is consistent with the variation in UV flux for a series of
spectra at maximum light, ranging from the broad, bright SN~1991T to the
fast, faint SN~1991bg, presented in Figure 1 of  Nugent {\em  et~al.}
(1996).  This figure also showed a correlation of absolute magnitude
with ratios of spectral features on either side of  the Ca~II H and K
absorption trough at 3800 \AA\ and with ratios of Si~II absorption
features at 5800 \AA\  and 6150 \AA.

Such multiple correlations with absolute magnitude can provide
alternative methods for calibrating the SN~Ia candle.  In the current
analysis, we use them as cross-checks for the width-luminosity
calibration when the data set is sufficiently complete.  For future data
sets they may provide better, or more accessible, primary methods of
magnitude calibration.

\subsection{Correlations with Host Galaxy Properties}

There have been some indications that the low luminosity members of the
single-parameter SN~Ia family are preferentially found in spheroidal
galaxies (Hamuy {\em et~al.} 1995) or that the more luminous SNe Ia
prefer late spirals (Branch, Romanishin, \& Baron 1996).  If correct, this suggests that
calibration within the SN~Ia family, whether by light curve shape,
color, or spectral features, is particularly important when comparing
SNe Ia from  a potentially evolving mix of host galaxy types.

\section{The High-Redshift Supernova Data Set}

\subsection{Discovery and Classification}

The seven supernovae discussed in this paper were discovered during
1992--94 in coordinated search programs at the Isaac Newton 2.5 m
telescope (INT) on La Palma and the Kitt Peak 4 m telescope, with
follow-up photometry and spectroscopy at multiple telescopes, including
the William Herschel 4 m, the Kitt Peak 2.3 m,  the Nordic Optical 2.5
m,  the Siding Springs 2.3 m, and the Keck 10 m telescopes. The
light-curve data were primarily obtained in the Johnson-Cousins $R$-band
(Harris set; see Massey {\em et~al.} 1996), with some additional data
points in the Mould $I$,  Mould $R$, and Harris $B$ bands. SN~1994G was
observed over the peak of the light curve in the Mould $I$ band  (Harris
set).   All of the supernovae were followed for more than a year past
maximum brightness so that the host galaxy light within the supernova
seeing disk could be measured and subtracted from the supernova
photometry measurements.   Spectra were obtained for each of the
supernova's host galaxies, and for the supernova itself in the case of
SN~1994F, SN~1994G, and SN~1994an. The redshifts were measured from
host-galaxy spectral features, and their uncertainties are all
$\ls0.001$. Table~\ref{tableone} lists the primary observational data
obtained from the light-curve photometry and spectroscopy.

\placetable{tableone}

We consider several types of observational evidence in classifying a
supernova as a type Ia:   (a)  Ideally, a spectrum of the supernova is
available and it matches the spectrum of a low-redshift type Ia
supernova observed  the appropriate number of rest-frame  days past its
light-curve maximum (e.g., Filippenko 1991.)  This will usually differentiate types Ib, Ic, and
II from type Ia.  For example, near maximum light SNe Ia usually develop  
strong, broad features, while the spectra of SNe II are more
featureless. Distinctive features such as a trough at  6150\AA\ (now
believed to be due to blueshifted Si~II $\lambda$6355) uniquely specify
a SN~Ia.   (b) The spectrum and morphology of the host galaxy can
identify it as an elliptical or S0, indicating that the supernovae is a
type Ia, since only SNe Ia are found in these galaxy types.  (Of course,
the converse does not hold, since SNe Ia are also found in late spirals,
at an even higher rate than in ellipticals, locally.)  We will here
consider E or S0 host galaxies to indicate a SN~Ia, with the caveat that
it is possible, in principle, that someday a SN~II may be found in these
galaxy types. (c) The light curve shape can narrow the range of possible
identifications by ruling out the plateau light curves of SNe IIP.  (d)
The statistics of the other classified supernovae discovered  in the
same search provide a probability that a random unclassified member of
the sample is a SN~Ia (given similar magnitudes above the detection
threshold).

Five of the seven supernovae discussed in this paper  can be classified
according to the criteria (a) and (b).  Two are confirmed to be SNe Ia
and three are consistent with SNe Ia:  The spectrum of SN~1994an 
exhibits the major spectral features from 3700\AA\ to 6500\AA (SN rest
frame wavelength) characteristic of SNe Ia $\sim$3 days past maximum
(rest frame time), including the   Si~II absorption near 6150\AA.    SN
1994am  was discovered in an elliptical galaxy, as identified by its
morphology in a Hubble Space Telescope WFPC2 image
(Figure~\ref{morphology}), and by  its spectrum, which matches that of a
typical present-day E or S0 galaxy. We take these two SNe, 1994an and
1994am, to be SNe Ia. The spectrum of the SN~1994al  host galaxy  is a
similarly good match to an E or S0, but until the morphology can be
confirmed we consider SN~1994al to be {\em consistent} with a  SN~Ia.

\begin{figure}[tbp]
\psfig{figure=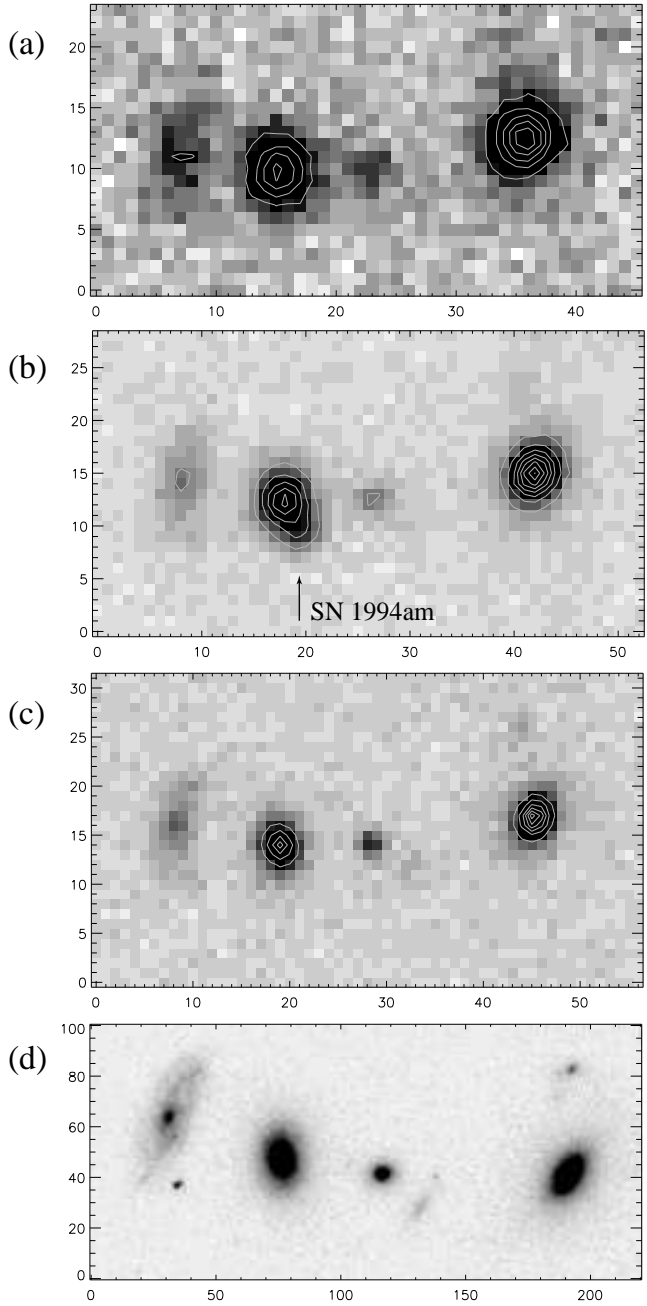,height=6.8in}
\caption[lvsm]{
Enlarged subimage showing the host galaxy
of SN~1994am, and three neighboring galaxies before, during, and after
the supernova explosion. (a)~Isaac Newton 2.5-m Harris $R$-band
observation on  19 December 1993, with 0.6 arcsec/pixel.   (b)~Kitt Peak
4-m Harris $R$-band observation on 4 February 1994, with 0.47
arcsec/pixel,  near maximum light of the supernova (indicated by arrow).
(c)~Cerro Tololo 4-m Harris $R$-band observation on 15 October 1995,
with 0.47 arcsec/pixel. (d)~Hubble Space Telescope (WFPC2) F814W
$I$-band observation on 13 January 1995, with 0.1 arcsec/pixel.  At this
resolution, the host  galaxy of SN~1994am can be identified as an
elliptical, providing strong evidence that  SN~1994am is a type Ia
supernova.}
\label{morphology}
\end{figure}

Spectra of both SN~1994F and SN~1994G  are more consistent with an SN~Ia
spectrum at the appropriate number of rest-frame days ($t_{\rm spect}$
of Table~\ref{tableone}) past the light curve maximum than any
alternatives.   The
spectrum of  SN~1994G  was observed $\sim $13 days past $B$ maximum
(rest frame) at the MMT by  P. Challis, A. Riess, and R. Kirshner and
$\sim$15 days past maximum at the Keck telescope.   The strengths of the
Ca~II H and K,  Fe, and Mg II features closely match those of a SN~Ia.  
The spectrum of  SN~1994F  was observed $\sim$2 days before maximum 
(rest frame) by  J.~B. Oke, J. Cohen, and T. Bida during the 
commissioning of LRIS at the Keck telescope, and therefore was neither
calibrated nor optimally  sky-subtracted.  The stronger SN~I  features
(e.g. Ca~II) do appear to be present nonetheless, so SN~1994F is
unlikely to be a luminous SN~II near maximum. The spectra for all of
these supernovae will be reanalyzed and published once the late-time
host-galaxy spectra are available, since host-galaxy features can
confuse details of the supernova spectrum.  
  
To address the remaining two unclassified supernovae, we consider the
classification statistics of our entire sample of  $>$28 supernovae
discovered to date by the Supernova Cosmology Project.  Eighteen
supernovae have been observed spectroscopically, primarily at the Keck
10 m telescope.  Of these, the 16 supernovae at redshifts $z \ge 0.35$
discovered by our standard search technique  are all consistent with a
type Ia identification, while the two closer ($z< 0.35$) supernovae are
type II.   This suggests that $>$94\% of the supernovae discovered by
this search technique at redshifts $z \ge 0.35$ are type Ia.  Moreover,
the shapes of the light curves we observe are inconsistent with the
``plateau'' light curves of  SNe IIP, further reducing the probability
of a non-type Ia. In this paper, we will therefore assume an SN~Ia
identification  for the remaining two unclassified supernovae.  

\subsection{Photometry Reduction}

The two primary stages of the photometry reduction are the measurement
of the supernova flux on each observed point of the light curve, and the
fitting of these data to a SN~Ia template to obtain peak magnitude and
width (stretch factor). At high redshifts, a significant fraction
($\sim$50\%) of the  light from the supernova's host galaxy usually lies
within the seeing-disk of the supernova, when observed from ground-based
telescopes. We subtract this host galaxy light from each photometry
point on the light curve.  The amount of host galaxy light to be
subtracted from a given night's observation is found from an aperture
matched to the size and shape of the point-spread-function for that
observation, and measured on the late-time images that are observed
after the supernova has faded. 

The transmission ratio between the late-time image and each of the other
images on the light curve is calculated from the objects neighboring the
supernova's host galaxy that share a similar color (typically $>$25
objects are used).   This provides a ratio that is suited to the
subtraction of the host galaxy light. Another ratio between the images
is calculated for the supernova itself, taking into account the
difference between the color of the host galaxy and the color of the
supernova at its particular time in the light curve by integrating
host-galaxy spectra and template supernova spectra over the filter and
detector response functions.  (These color corrections were not always
necessary since the filter-and-detector response function often matched
for different observations.) The magnitudes are thus all referred to the
late-time image, for which we observe a series of Landolt (1992)
standard fields  and globular cluster tertiary standard fields
(Christian {\em et al.} 1985; L. Davis, private communication). The
instrumental color corrections account for the small differences between
the instruments used, and between the instruments and the Landolt
standard filter curves. 

This procedure is checked for each image by similarly subtracting the
light in apertures on numerous ($\sim$50)  neighboring galaxies of 
angular size, brightness, and color similar to the host galaxy.  If
these test apertures, which contain no supernova, show flat light curves
then the subtraction is perfect.  The RMS deviations from flat
test-lightcurves provide an approximate measurement,  $\sigma_{\rm
test}$, of photometry uncertainties due to the matching procedure
together with the expected RMS scatter due to photon noise.  We 
generally reject images for which these matching-plus-photon-noise
errors are more than 20\% above the  estimated photon-noise error alone
(see below). (Planned HST light curve measurements will make these steps
and checks practically unnecessary, since the HST  point-spread-function
is quite stable over time, and small enough that the host galaxy will
not contribute substantial light  to the SN~measurement.)

{\em Photometry Point Error Budget.} 
We track the sources of photometric uncertainty at each step of this
analysis to construct an error budget for each photometry point of each
supernova. The test-lightcurve error, $\sigma_{\rm test}$, then provides
a check of  almost all of these uncertainties combined. The dominant
source of photometry error in this budget  is the Poisson fluctuation of
sky background light, within the photometry apertures, from the
mean-neighborhood-sky level.  A typical mean sky level on these images
is $N_{\rm sky} = 10,000$ photoelectrons (p.e.) per pixel, and it is
measured to approximately 2\% precision from the neighboring region on
the image.  Within a 25-pixel aperture, the sky light contributes a
Poisson noise of ($25 N_{\rm sky})^{1/2} \approx 500$ p.e. The light in
the aperture from the supernova itself and its host galaxy  is typically
$\sim$6,000 p.e., and contributes only $\sim$77 p.e. of noise in
quadrature,  negligible compared to the sky background noise. Similarly,
the noise contribution from the sky light in the subtracted late-time
images after the supernova fades is also negligible, since the late time
images are typically $\gs$9 times longer exposures then the other
images. The supernova photometry points thus typically each have 
$\sigma_{\rm photon} \approx $ 11\% photon-noise measurement error.

Much smaller contributions to the error budget are added by the   
previously-mentioned calibration and correction steps: The magnitude
calibrations have uncertainties between 1\% and 4\%, and uncertainties
from instrumental color corrections  are  $\leq$1\%.  Uncertainties
introduced in the ``flat-fielding'' correction for pixel-to-pixel
variation in quantum efficiency are less than  $1/\sqrt 10$ of the sky
noise per pixel, since $\gs$10 images are used to calculate the ``flat
field'' for quantum efficiency correction.  This flat-fielding error
leads to $\sim$32 p.e. uncertainty in the supernova-and-host-galaxy
light, which is an additional 5\% uncertainty in quadrature above
$\sigma_{\rm photon}$. For most images the quadrature sum of  all of
these error contributions agrees with the overall test dispersion,
$\sigma_{\rm test}$.  The exceptions mentioned above  generally arise
from point-spread-function variations over a given image in combination
with extremely poor seeing, and such images are generally rejected
unless  $\sigma_{\rm test}$ is within 20\% of the expected quadrature
sum of the error contributions.

Since the target fields for the supernova search were chosen at high
Galactic latitudes whenever possible, the uncertainties in our Galaxy
extinction,  $A_R$ (based on values of $E(B-V)$ from Burstein \& Heiles
1982), are generally $<$1\% on the photometry measurement.   The one
major exception is SN~1994al,  for which there is more substantial
Galactic extinction, $A_R = 0.23$~mag, and hence we quote a more
conservative extinction error of $\pm$0.11 mag for this supernova.

All of these sources of uncertainties are included in the error budget
of Table~\ref{tableone}. 
The photometry data points derived from this first stage of
the reduction are shown in Figure~\ref{lightcurves}. Although we observe
$\ge$2 images for each night's data point to correct for cosmic rays and
pixel flaws, these images have been combined in producing these plots
(and in the preceding noise calculations, for direct comparison). Since
the error bars depend on the uncertainty in the host-galaxy measurement,
there is significant correlation between them, particularly for nights
with similar seeing.   Therefore, the usual point-to-point statistics
for uncorrelated data do not apply for these plots. Further details
concerning the calibration observations, color corrections, and data
reduction, with mention of specific nights, telescopes, and supernovae,
are given in  Perlmutter {\em et~al.} (1995), and the forthcoming data
catalog paper.

\begin{figure}[tbp]
\psfig{figure=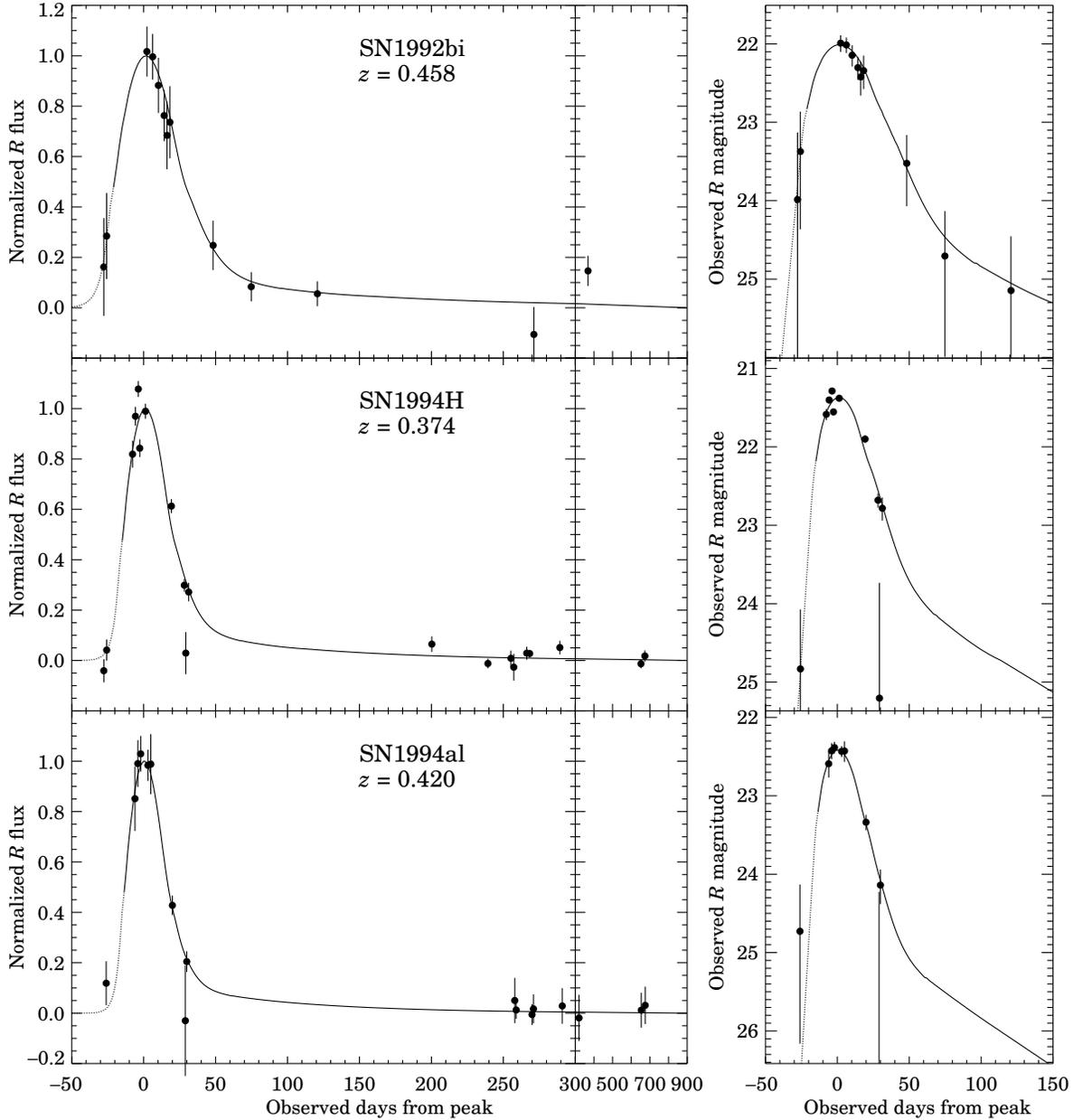,height=6.5in}
\caption[lvsm]{
{\em (Continued on next page.)}  $R$-band
light curve photometry points for the first seven high-redshift
supernovae discovered by the Supernova Cosmology Project, and best-fit
template SN~Ia light curves.  The left panels show the relative flux as
function of observed time (not supernova-rest-frame time).  The right
panels show observed $R$-band magnitude versus observed time. Note that
there is significant correlation between the error bars shown,
particularly for observations with similar seeing, since the error bars
depend on the uncertainty in the host-galaxy measurement that have been
subtracted from these measurements (see text). An $I$-band light curve
is also shown for SN~1994G; other photometry points in  $I$ and $B$ for
the seven supernovae are not shown on this plot. The rising slope (in
mag/day) of the template light curve before rest-frame day $-$10
(indicated by the grey part of the curves) is not well-determined, since
few low-redshift supernovae are discovered this soon before maximum
light.  A range of possible rise times was therefore explored (see
text).}
\label{lightcurves}
\end{figure}
\addtocounter{figure}{-1}
\begin{figure}[tbp]
\psfig{figure=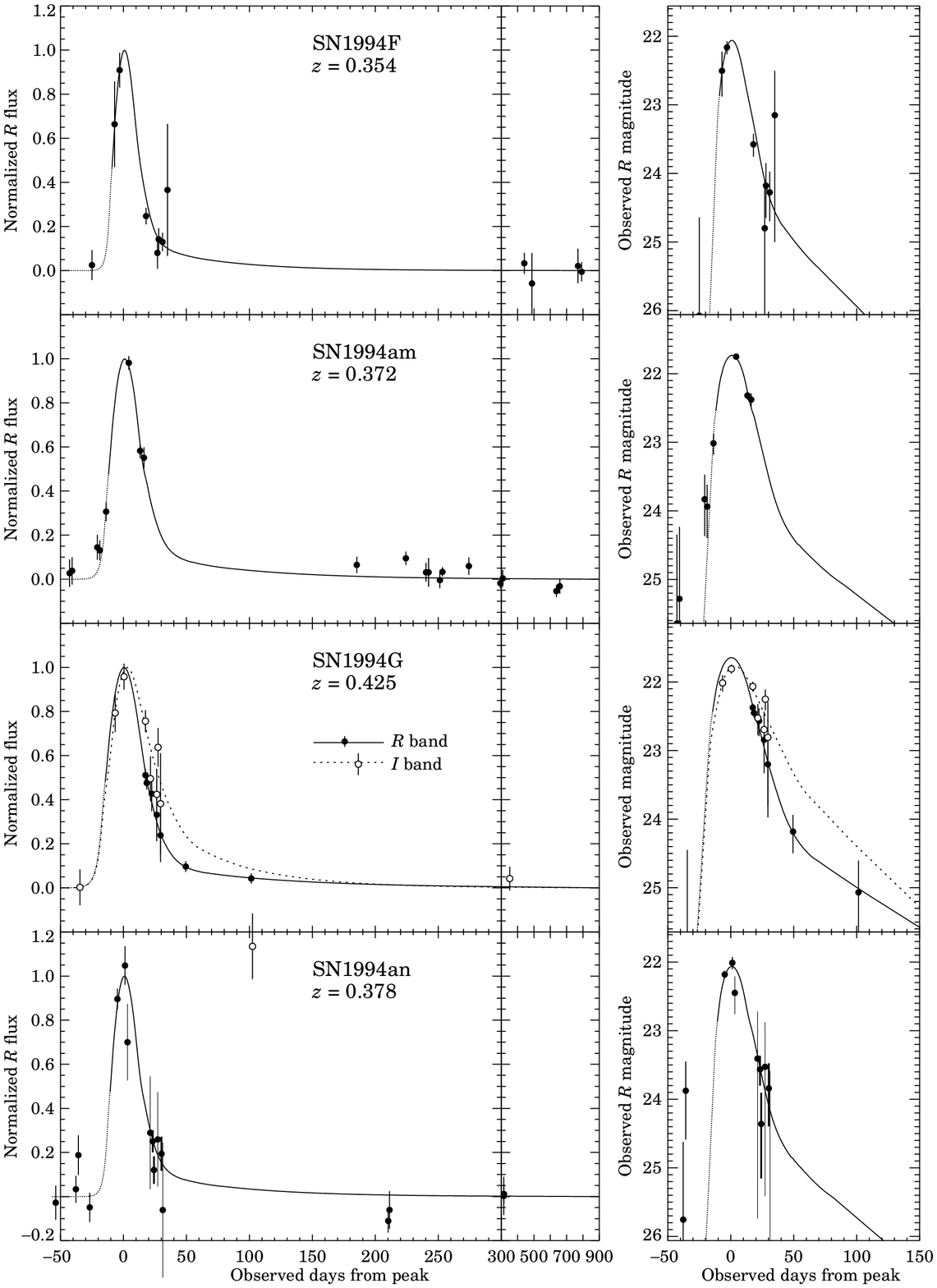,height=8.4in}
\caption[lbsm]{
{\em continued}.}
\label{lightcurves2}
\end{figure}


\subsection{Light Curve Fit}

At $z\approx 0.4$, the light that leaves the supernova in the $B$ band
arrives at our telescopes approximately in the $R$ band. The second
stage of the photometry reduction, fitting the light curve, must
therefore be performed using a $K$-corrected SN~Ia light curve template:
We use spectra of several low-redshift supernovae,  well-observed over
the course of their light curves, to calculate a table of cross-filter
$K$ corrections, $K_{BR}(t)$, as a function of light curve time.  These
corrections account for the mismatch between the redshifted $B$ band and
the $R$ band (including the stretch of the  transmission-function
width), as well as for the difference in the  defined zero points of the
two magnitude systems.  For each high-redshift supernova, we can then
construct a predicted $R$-band template light curve as would be observed
for the redshift, ${\cal T}_R^{\rm observe}$, based on the $B$-band
``standard'' template (i.e., with $\Delta m_{15}=1.1$~mag)  as actually
observed in the SN rest frame:
\begin{equation}
  {\cal T}_R^{\rm observe}(t') =  {\cal T}_B^{\rm restframe}(t) + K_{BR}(t),
\label{templates}
\end{equation}
where the observed time dependence, $t' \equiv t(1+z)$, accounts for the
time dilation of events at redshift $z$. The calculations of $K_{BR}(t)$
are described and tabulated, in Kim, Goobar, \& Perlmutter (1996), with
an error analysis that yields uncertainties of $<$0.04 mag for redshifts
$z< 0.6$.  Table~1 lists the values of $K_{BR}$ at maximum light for the
redshifts of the seven supernovae.

For the exceptional case of SN~1994al, with significant $E(B$$-$$V)$
from our own Galaxy, we also calculate the extinction, $A_R$, as a
function of supernova-rest-frame time, again using a series of 
redshifted supernova spectra multiplied by a reddening curve for an
$E(B$$-$$V)_{\rm B\&H}$ value given by Burstein \& Heiles (1982).  For
this particular supernova's redshift,  $A_R=0.23 \pm 0.11$,  with
variations of  0.01 magnitude for the dates observed on the light curve.  
For the redshifts of all seven supernovae, we find essentially the same
ratio  ${\cal R}_R = A_R /  E(B$$- $$V)_{\rm B\&H} = 2.58 \pm 0.01$, at
maximum light.

The photometry data points for each supernova are fit to the as-observed
$R$-band template, ${\cal T}_R^{\rm observe}$, with free parameters for
the stretch-factor, $s$, the  $R$ magnitude at peak, $m_R$, the date of
peak, $t_{\rm peak}$, and the  additive constant flux, $g_{\rm resid}$,
that accounts for the residual host-galaxy light due to Poisson error in
the late-time image:
\begin{equation} 
   f(t;m_R,s,t_{\rm peak}, g_{\rm resid}) = 
   10^{-0.4\left[m_R-m_R^{\rm 0} + 
   {\cal T}_R^{\rm observe}(t's-t_{\rm peak})\right]}+ g_{\rm resid} \;,
\label{fitfunc}
\end{equation}
where $m_R^{\rm 0}$ is the calibration zero-point for the $R$
observations and ${\cal T}_R^{\rm observe}(t)$ is normalized to zero at
$t=0$. (For one of the supernovae with sufficient $I$ band data,  SN
1994G, the fitting function $f$ also includes the redshifted,
$K$-corrected $V$-band template, and an additional parameter for the
$R$$-$$I$ color at maximum is fit.) We fit to flux measurements, rather
than magnitudes, because the error bars are symmetric in flux, and
because the data points have the late-time galaxy light subtracted out
and hence can be negative.  In this fit, particular care is taken in
constructing the covariance matrix  (see, e.g., Barnett  {\em et~al.}
1996) to account for the correlated photometry error due to the fact
that the same  late-time images of the host galaxy are used for all
points on a light curve. 

{\em Light Curve Error Budget.}
We compute uncertainties for  $m_R$ and $s$  using both a Monte Carlo
study and a mapping of the $\chi^2$ function; both methods yield similar
results.   Table~\ref{tableone} lists the best-fit values and uncertainties for 
$m_R$ and $s$.  Usually, many data points contribute to the template
fit, so the uncertainty in the peak, $m_R$, is less than the typical
$\sim$11\% photometry uncertainty in each individual point.  However,
our Monte Carlo studies show that it is usually very important that high
quality late-time and pre-maximum data points be available to constrain
both $s$ and $m_R$.  We have explored the consequences of adding or
subtracting observations on the light curves, although the error bars
reflect the actual light curve sampling observed for a given  supernova. 

The rising slope of the template light curve before rest-frame day $-$10
(indicated by the grey part of the curves in Figure~\ref{lightcurves})
is not well-determined, since few low-redshift supernovae are discovered
this soon before maximum light.  A range of possible rise times was
therefore also explored.  Only two of the supernovae show any
sensitivity to the choice of rise time.  The effect is well within the
error bars of the stretch factor, and negligible for the other
parameters of the fit.

For the analyses of this paper, we translate these observed $R$ magnitudes
back to the ``effective'' $B$ magnitudes,  $m_B = m_R - A_R - K_{BR}$,
where all the quantities (see Table~\ref{tableone}) are for the light curve
$B$-band peak (SN rest frame), and we have corrected for Galactic
extinction, $A_R$, at this stage.  In the following section, we directly
compare these ``effective'' $B$ magnitudes  with the $B$ magnitudes of 
the low-redshift supernovae using the equations of Section 2,
substituting the $K$-corrected effective $B$ magnitudes, $m_B$,  and the
$B$-band zero point, ${\cal M}_B$, for the bolometric magnitudes, $m$
and ${\cal M}$.

\section{Results for the High-Redshift Supernovae}

\subsection{Dispersion and Width-Brightness Relation}

Before using a width-luminosity correction, it is  important to test
that it applies at high redshifts, and that the magnitude dispersions
with and without this correction are consistent with those of the
low-redshift supernovae. We study the peak-magnitude dispersion of the
seven high-redshift supernovae by calculating their absolute magnitudes
for an arbitrary choice of  cosmology.   This allows the relative
magnitudes of the supernovae at  somewhat different redshifts ($z =
0.35$--0.46) to be compared.  The slight dependence on the choice of
cosmology is negligible for this purpose, for a wide range of
$\Omega_{\rm M}$ and   $\Omega_\Lambda$.   Choosing $\Omega_{\rm M} = 1$
and $\Omega_\Lambda=0$,  the RMS dispersion about the mean absolute
magnitude  for the seven supernovae is $\sigma_{M_B} = 0.27$. (We find
the same RMS dispersion for the best-fit cosmology discussed below in
Section 5.3.)

\begin{figure}[tbp]
\psfig{figure=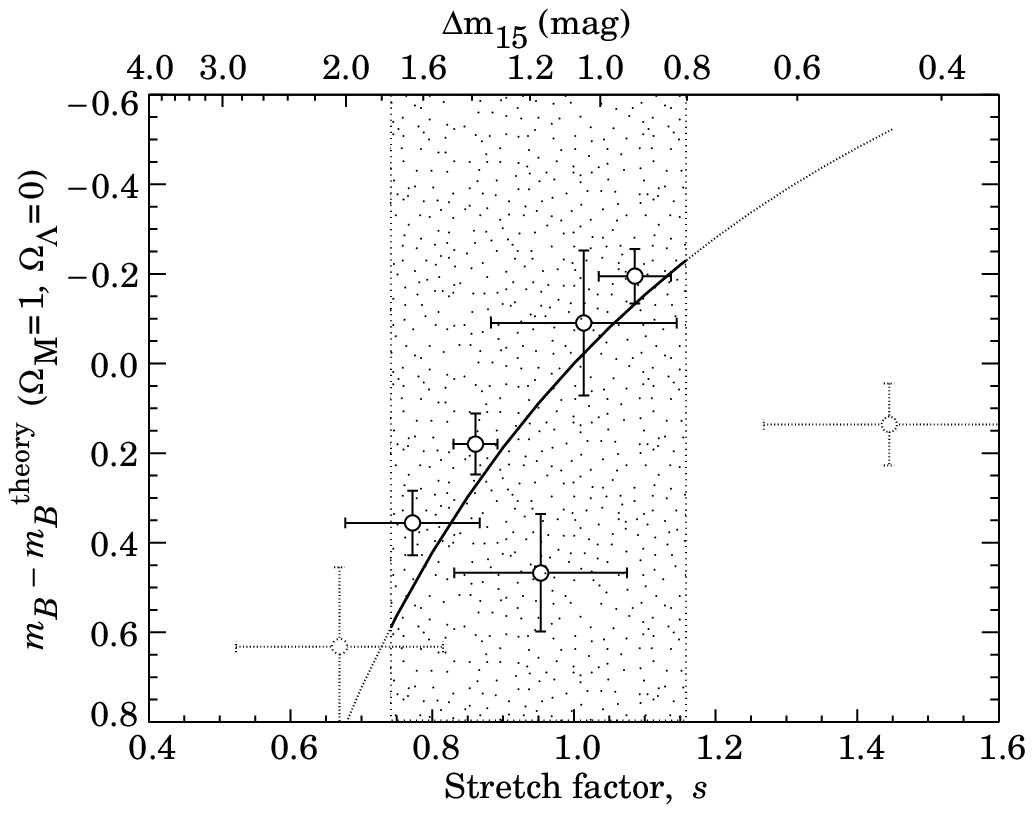,height=4.2in}
\caption[lvsm]{
The difference between the measured
and ``theoretical'' (Equation~\ref{simplemagz}   for $\Omega_{\rm M} =
1$,  $\Omega_\Lambda=0$, and ${\cal M} = {\cal M}_{B,{\rm
corr}}^{\scriptscriptstyle\{1.1\}}$)  $B$ magnitudes (uncorrected for
the width-luminosity relation) versus the best-fit stretch factor, $s$,
for the high-redshift supernovae.  The stretch factor is fit in the
supernova rest frame, i.e., after correcting for the cosmological time
dilation calculated from the host-galaxy redshift  (see
Equations~\ref{templates} and \ref{fitfunc}). If different values of
($\Omega_{\rm M}$, $\Omega_\Lambda$) had been chosen the labels would
change, but the data points would not vary significantly within their
error bars, since the range of redshifts is not large for these
supernovae. The upper axis gives the equivalent values of $\Delta m_{15}
= 1.96(s^{-1}-1) + 1.07$ (Eq.~\ref{stodeltam15}). The solid line shows
the width-luminosity relation (Eq.~\ref{widthbrightrel})  found by Hamuy
{\em et~al.} (1995, 1996) for an independent set of 18 nearby ($z \le
0.1$) SNe Ia, for which $0.8 \ls \Delta m_{15} \ls  1.75$~mag. This
range of light-curve widths is indicated by the shaded region.   The
curve and data points outside of this range are plotted in a different
shade to emphasize that the relation is only established within this
range. A different choice of  ($\Omega_{\rm M}$, $\Omega_\Lambda$) or of
the magnitude zero-point ${\cal M}$ would move the line in the vertical
direction relative to the points.}
\label{broadbright}
\end{figure}

Figure~\ref{broadbright} shows the difference between the measured and
the ``theoretical'' (Equation~\ref{simplemagz} for $\Omega_{\rm M} = 1$,  
$\Omega_\Lambda=0$ and ${\cal M} = {\cal M}_{B,{\rm
corr}}^{\scriptscriptstyle\{1.1\}}$)  uncorrected effective $B$
magnitudes,  $m_B -m_B^{\rm theory}$, as a function of the best-fit
stretch factor $s$ or, equivalently, $\Delta m_{15}$.  The zero-point
intercept, ${\cal M}_{B,{\rm corr}}^{\scriptscriptstyle\{1.1\}}$, from 
Equation~\ref{widthbrightrel} was used in the calculation of  $m_B^{\rm
theory}$, so that the data points could be compared to the
Equation~\ref{widthbrightrel} width-luminosity relation  found for the
nearby sample of Hamuy {\em et~al.} (the solid line of
Figure~\ref{broadbright}).   A different choice of  ($\Omega_{\rm M}$,
$\Omega_\Lambda$) or of the magnitude zero-point ${\cal M}$ would move
the line in the vertical direction relative to the points,  so the
comparison should be made with the shape of the relation, not the exact
fit.  The width-luminosity relation seen for nearby supernovae appears
qualitatively to hold for the high-redshift supernovae. 

To make this result quantitative, we use the prescription of 
Equation~\ref{widthbrightrel} to ``correct" the magnitudes  to that of a
supernova with $\Delta m_{15} = 1.1$~mag, by adding magnitude
corrections, $\Delta_{\rm corr}^{\scriptscriptstyle\{1.1\}} = -0.86 
(\Delta m_{15} - 1.1)$, where we derive $\Delta m_{15}$ from $s$ using
Equation~\ref{stodeltam15}. Note that the resulting corrected magnitude,
$m_{B,{\rm corr}} = m_B + \Delta_{\rm
corr}^{\scriptscriptstyle\{1.1\}}$, listed in  Table~\ref{tableone}, has an
uncertainty that is not necessarily the quadratic sum of the
uncertainties for $m_B$ and $\Delta_{\rm
corr}^{\scriptscriptstyle\{1.1\}}$,  because the best-fit values for $s$
and  $m_R$ (and hence $m_B$) are correlated in the light-curve template
fit.

Two of the high-redshift supernovae (shown with different-shaded symbols
in Figure~\ref{broadbright} and subsequent figures)  have values of
$\Delta m_{15}$ that are outside of the range of values studied in the
low-redshift supernova data set  of Hamuy {\em et~al}. The
width-luminosity correction for these supernovae is therefore less
reliable than for the other five supernovae, since it depends on an
extrapolation. It is also possible that the true values of $\Delta
m_{15}$ for the two supernovae fall within this range, given the fit
error bars.     (One of these, SN~1992bi at $z=0.458$, has very large
error bars because its light curve has larger photometric uncertainties
and the  lightcurve sampling was not optimal to constrain the stretch
factor $s$.)   In Table~\ref{tableone}, the corrections, $\Delta_{\rm
corr}^{\scriptscriptstyle\{1.1\}}$, and corrected magnitudes, $m_{B,{\rm
corr}}$,  for these supernovae are therefore listed in brackets, and we
also give the corrected magnitudes,  $m_{B,{\rm corr}}^{\rm extreme}$,
obtained using only the most extreme   $\Delta_{\rm
corr}^{\scriptscriptstyle\{1.1\}}$ corrections found for low-redshift
supernovae. In the following analysis we calculate all results using
just the other five supernovae that are within the Hamuy {\em et~al.}
range of  $\Delta m_{15}$. As a cross-check, we then provide the result
using all seven supernovae,  but without width-luminosity correction.

The peak-magnitude RMS dispersion for the five high-redshift supernovae
that are within the $\Delta m_{15}$ range  improves from $\sigma_{M_B} =
0.25$~mag before applying the  width-luminosity correction to
$\sigma_{M_B,{\rm corr}} = 0.19$~mag  after applying the correction. 
These values agree  with those for the 18 low-redshift Cal\'{a}n/Tololo
supernovae,  $\sigma_{M_B}^{\rm Hamuy} = 0.26$   and $\sigma_{M_B,{\rm
corr}}^{\rm Hamuy} = 0.17$~mag, suggesting that the supernovae at
high-redshift are drawn from a similar population, with a similar
width-luminosity correlation.   Note that although the slope of  the
width-luminosity relation, Equation~\ref{widthbrightrel}, stays the
same, it is possible that the intercept could change at high redshift,
but this would be a remarkable conspiracy of physical effects, since the
time scale of the event appears naturally correlated with the strength
and temperature of the explosion (see Nugent {\em et al.} 1996).

\subsection{Color and Spectral Indicators of Intrinsic Brightness}

With the supplementary color information in the $I$ and $B$ bands, and
the spectral information for three of the supernovae, we can also begin
to test the other indicators of intrinsic brightness within the SN~Ia
family. The broadest, most luminous supernova of our five-supernova
subsample,  SN~1994H,  is bluer than $B$$-$$R=0.7$ at the 95\%
confidence level $2\pm 2$ days before maximum light (this is only a
limit because of possible clouds on the night of the $B$ calibration).
For comparison, Nugent (1996, private communication) has estimated
observed $B$$-$$R$ colors within $\sim$4 days of  maximum light as a
function of redshift using spectra for a range of SN~Ia sub-types, from
the broad, superluminous SN~1991T to the narrow, subluminous SN~1991bg.  
SN~1994H is only slightly bluer than the  $B$$-$$R \approx 1.0 \pm 0.15$
color of SN~1991T at $z=0.374$ (where the error represents the  range of
possible host-galaxy reddening).  However,  SN~1994H does not agree with
the redder colors (at $z=0.374$) of  SN~1981B ($B$$-$$R \approx 1.7$
mag) and SN~1991bg ($B$$-$$R \approx 2.6$~mag).   SN1991T has a broad
light curve width of  $\Delta m_{15}=0.94 \pm 0.07$, which agrees within
error with  the light curve width of SN~1994H,   $\Delta m_{15}=0.91 \pm
0.09$, whereas  SN~1981B and SN~1991bg  both have ``normal'' or narrower
light curve widths, $\Delta m_{15}=1.10 \pm 0.07$ and $\Delta
m_{15}=1.93 \pm 0.10$~mag.

The multi-band version of the ``correction template'' analysis  (Riess
{\em et al.} 1996) is designed to fit simultaneously for the host galaxy
extinction (discussed later) and the intrinsic brightness of the
supernova. The rest frame $V$ light curve (approximately redshifted into
our observed $I$) is the strongest indicator of the SN~Ia family
parameterization when using this approach. We thus use this technique to
analyze SN1994G, with its well sampled $I$ light curve in addition to
$R$ data.   We find that the supernova is best fit by a correction
template that indicates that it is intrinsically overluminous by $0.06
\pm 0.38$ magnitudes, compared to a Leibundgut-template supernova. This
is consistent with the $0.05 \pm 0.22$ overluminosity found using  the
width-luminosity correlation and the best-fit stretch factor.

(The ``correction template'' fit  gives larger error bars than a simple
stretch-factor fit because the correction templates have significant
uncertainties, with day-to-day correlations that are, unfortunately, not
tabulated.   Currently, the fits to $B$-band correction templates have
even larger uncertainties than $V$-band, possibly due to a poor fit of
the linear correction model to the low-redshift data.  This approach is
therefore not useful for the other six high-redshift supernovae of this
first set, since  these large correction-template uncertainties 
propagate into $>$1 magnitude uncertainties in the fits to the observed
$R$ band data.)

Both intrinsically fainter supernovae and  supernovae with host galaxy
extinction can appear redder in $R$$-$$I$ (corresponding to
approximately $B$$-$$V$ in the supernova rest frame).  For the several
supernova for which we  have scattered $I$-band photometry, we thus can
use it to  estimate extinction only after the supernova sub-type has
been determined using the stretch factor. SN~1994G   fit closely to the
$s=1$ template.  An $s=1$ supernova at $z=0.420$ will have an expected
observed color $R-I =0.16 \pm 0.05$~mag  at observed-$R$ maximum light.
For SN~1994G, we observed $R-I=-0.12 \pm 0.17$~mag.  This gives
$E(R$$-$$I)_{\rm observed} \approx E(B-V)_{\rm restframe}  = -0.28 \pm
0.18$~mag, or $A_V < 0.01$~mag at the 95\% confidence level. The Riess
{\em et al.} ``correction template'' analysis of SN~1994G  also yields a
bound on extinction of  $A_V<0.01$~mag (90\% confidence).  

The more recent high-redshift supernovae studied by the Supernovae
Cosmology Project have more complete color and spectral data, and so
should soon provide further tests of these  additional SN~Ia luminosity
indicators. In addition, the spectrum of SN~1994an covers both
wavelength regions in which the Nugent {\em et~al.} (1996) line ratios
correlate well with supernova magnitudes. This analysis awaits the
availability of the spectrum of the host galaxy without the supernova
light, so that we can subtract the galaxy-spectrum ``contamination''
from the supernova spectrum.  Further late-time $B$-band images of SN
1994an  will also allow $B$$-$$R$ color cross-checks using $B$ images
that were observed four days after maximum.

\subsection{Magnitude-Redshift Relation and the Cosmological Parameters}

\begin{figure}[tbp]
\psfig{figure=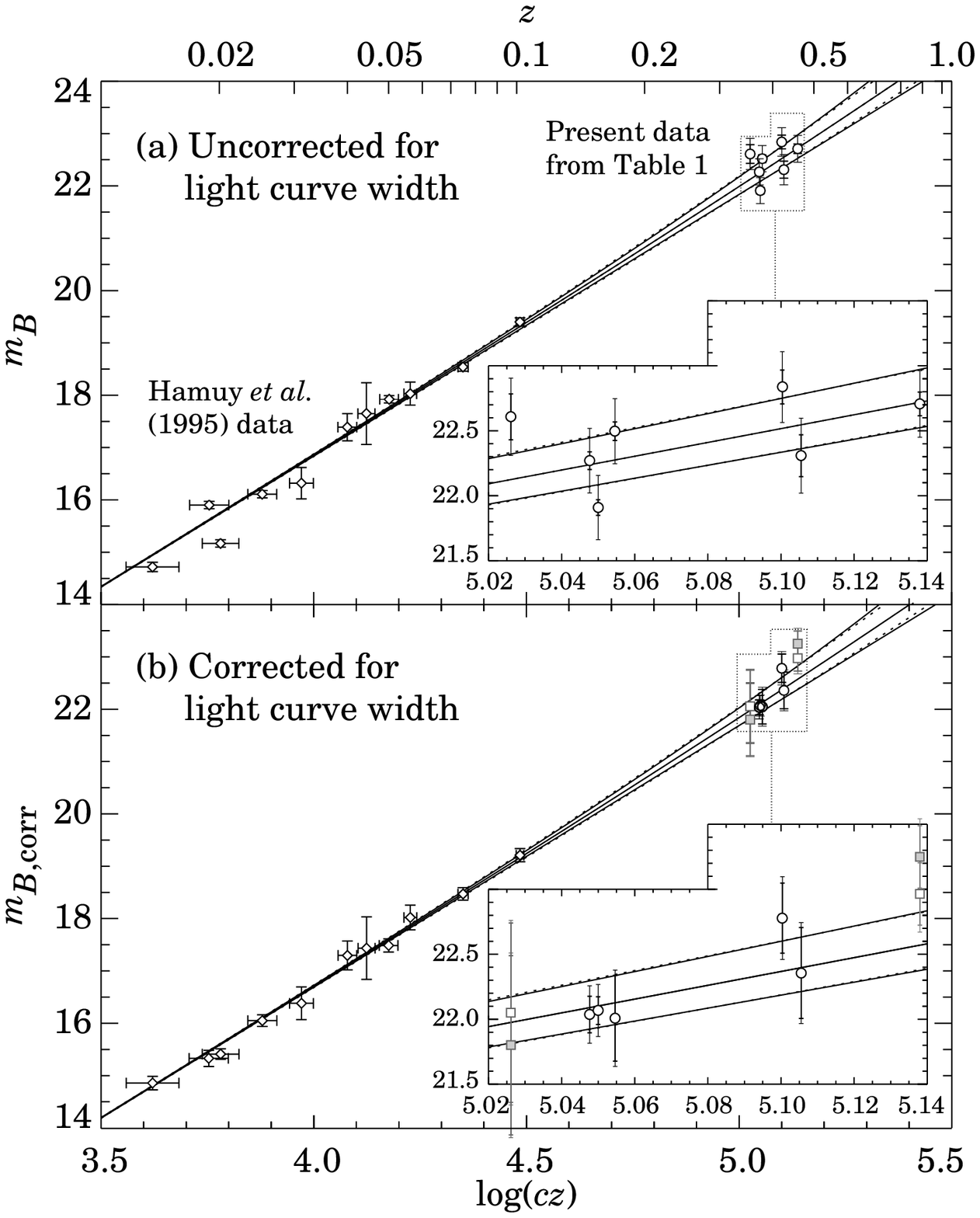,height=6.5in}
\caption[lvsm]{
Hubble diagrams for the first seven
high-redshift supernovae, (a) uncorrected $m_B$, with low-redshift
supernovae of Hamuy {\em et~al.} (1995) for visual comparison;  (b)
$m_{B,{\rm corr}}$ after ``correction'' for width-luminosity  relation.
The square points are not used in the analysis, because they are
corrected based on an extrapolation outside the range of light curve
widths of low-redshift supernovae (see text and Table~\ref{tableone}). Insets
show the high-redshift supernovae on magnified scales. The solid curves
in (a) and (b) are theoretical $m_B$ for  ($\Omega_{\rm M}$,
$\Omega_\Lambda$) = (0, 0) on top, (1, 0) in middle, and (2, 0) on
bottom.  The dotted curves, which are practically indistinguishable from
the solid curves, are for the flat universe case, with ($\Omega_{\rm
M}$,  $\Omega_\Lambda$) =  (0.5, 0.5) on top, (1, 0) in middle, and
(1.5,~$-$0.5) on bottom. The inner error bars on the data points show
the photometry measurement uncertainty, while the outer error bars add
the intrinsic dispersions found for low-redshift supernovae,
$\sigma_{M_B}^{\rm Hamuy}$ for (a) and $\sigma_{M_B,{\rm corr}}^{\rm
Hamuy}$  for (b), for comparison to the theoretical curves.    Note that
the zero-point magnitude used for (b),  ${\cal M}_{B,{\rm
corr}}^{\scriptscriptstyle\{1.1\}}$,  and hence the effective $m_B$
scale,  is shifted slightly from the uncorrected ${\cal M}_{B}$ used for
the curves of (a).}
\label{magredshift}
\end{figure}

Figure~\ref{magredshift}(a) shows the Hubble diagram, $m_B$ versus $\log
cz$, for the seven high-redshift supernovae, along with low-redshift
supernovae of  Hamuy {\em et~al.} (1995) for visual comparison.  The
solid curves are plots of $m_B^{\rm theory}$, i.e.,
Equation~\ref{simplemagz}  with ${\cal M} = {\cal M}_{B}$ for  three
$\Lambda = 0$ cosmologies,  ($\Omega_{\rm M}$,  $\Omega_\Lambda$) = (0,
0), (1, 0), and (2, 0); the dotted curves, which are practically
indistinguishable from the solid curves, are for three flat cosmologies,
($\Omega_{\rm M}$,  $\Omega_\Lambda$) =  (0.5, 0.5), (1, 0), and
(1.5,~$-$0.5).  These curves show that the redshift range of the present
supernova sample  begins to distinguish the values of the cosmological
parameters.   Figure~\ref{magredshift}(b) shows the same
magnitude-redshift relation for the data after adding the
width-luminosity correction term, $\Delta_{\rm
corr}^{\scriptscriptstyle\{1.1\}}$; the curves of  $m_B^{\rm theory}$ in
(b) are calculated for ${\cal M} = {\cal M}_{B,{\rm
corr}}^{\scriptscriptstyle\{1.1\}}$.

\begin{figure}[tbp]
\psfig{figure=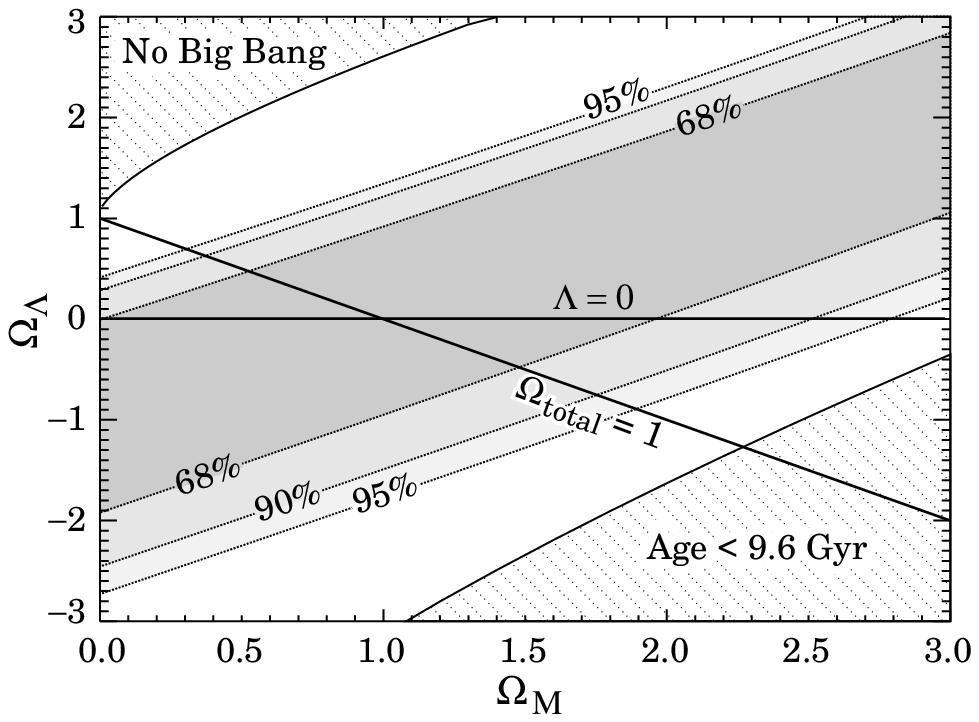,height=4.2in}
\caption[lvsm]{
Contour plot of the  68\% (1$\sigma$), 90\%, and
95\% (2$\sigma$) confidence regions in the $\Omega_{\Lambda}$ versus
$\Omega_{\rm M}$ plane, for the first seven high-redshift supernovae.
The solid lines show the two special cases of a $\Lambda = 0$ cosmology
and a flat ($\Omega_{\rm total} \equiv \Omega_{\rm M}  +\Omega_\Lambda =
1$) cosmology.   (Note that the constrained one-dimensional confidence
intervals for $\Lambda = 0$ or $\Omega_{\rm total} = 1$ are smaller than
the intersection of these lines with the two-dimensional contours, as
discussed in the text.)   The two labeled corners of the plot are ruled
out because they imply: (upper left corner) a ``bouncing''  universe
with no big bang (see Carroll {\em et~al.} 1992), or (lower right
corner) a universe  younger than the oldest heavy elements, $t_0 < 9.6$
Gyr (Schramm 1990), for any value of $H_0\ge 50$  km s$^{-1}$
Mpc$^{-1}$.}
\label{omegaomega}
\end{figure}

Figure~\ref{omegaomega} shows the  68\% (1$\sigma$), 90\%,  and 95\%
(2$\sigma$) confidence regions on the $\Omega_{\rm M}$--$\Omega_\Lambda$
plane for the fit of Equation~\ref{simplemagz} to the high-redshift
supernova magnitudes, $m_{B,{\rm corr}}$, after width-luminosity
correction, using  ${\cal M} = {\cal M}_{B,{\rm
corr}}^{\scriptscriptstyle\{1.1\}}$. In this fit, the magnitude
dispersion, $\sigma_{M_B,{\rm corr}}^{\rm Hamuy} = 0.17$, is added in
quadrature to the error bars of  $m_{B,{\rm corr}}$ to account for the
residual intrinsic dispersion, after width-correction, of our model 
(low-redshift) standard candles.

The two special cases represented by the solid lines of
Figure~\ref{omegaomega} yield significant measurements:   We again fit
Equation~\ref{simplemagz} to the high-redshift supernova magnitudes,
$m_{B,{\rm corr}}$, but this time with the fit constrained first to a
$\Lambda = 0$ cosmology (the horizontal line of 
Figure~\ref{omegaomega}), and then to a flat universe (the diagonal line
of  Figure~\ref{omegaomega}, with $\Omega_{\rm total} \equiv \Omega_{\rm
M} +\Omega_\Lambda = 1$). In the case of a $\Lambda = 0$ cosmology, we
find the mass density of the universe to be $\Omega_{\rm M} = 0.88
\;^{+0.69}_{-0.60}$.   For a flat universe (the diagonal line of 
Figure~\ref{omegaomega}, with $\Omega_{\rm total} \equiv \Omega_{\rm M}
+\Omega_\Lambda = 1$),  we find the cosmological constant to be
$\Omega_\Lambda = 0.06 \;^{+0.28}_{-0.34}$,  consistent with no
cosmological constant.   For comparison with other results in the
literature, we can also  express this fit as a 95\% confidence level
upper limit of  $\Omega_\Lambda < 0.51$.  Finally, this fit can be
described by the value of the mass density, $\Omega_{\rm M} = 1- 
\Omega_\Lambda =  0.94 \;^{+0.34}_{-0.28}$. (For brevity, we will
henceforth only quote the $\Omega_\Lambda$ fit results for the
flat-universe case, since either  $\Omega_\Lambda$ or $\Omega_{\rm M}$  
can be used to parameterize the flat-universe fit.) The goodness-of-fit
is quantified by $Q(\chi^2|\nu)$, the probability of obtaining the
best-fit $\chi^2$ or higher for  $\nu = N_{\rm SNe} -1$ degrees of
freedom  (following the notation of Press {\em et~al.} 1986, p. 165).
For both the $\Lambda=0$ and flat universe cases, $Q(\chi^2|\nu)= 0.75$.

The error bars on the measurement of  $\Omega_\Lambda$ or $\Omega_{\rm
M}$ for a flat universe are about two times smaller than the error  bars
on $\Omega_{\rm M}$ for a $\Lambda = 0 $ universe.   This can be seen
graphically in Figure~\ref{omegaomega}, in which the confidence region
strip    makes a shallow angle with respect to the $\Omega_\Lambda = 0$
line but crosses  the  $\Omega_{\rm total} = 1$ line at a sharper angle. 
Note that these error bars in the constrained one-dimensional fits for
$\Lambda = 0$ or $\Omega_{\rm total} = 1$ are smaller than the
intersection of the constraint lines of Figure~\ref{omegaomega} and the 
68\% contour band.  This is because a different range of  $\Delta\chi^2
\equiv \chi^2 - \chi^2_{\rm min}$ corresponds to 68\% confidence for one
free parameter, $\nu=1$ degree of freedom, as opposed to  two free
parameters, $\nu=2$ (see Press {\em et~al.} 1986, pp. 532-7).

We also analyzed the data for the same five supernovae ``as measured,''
that is, without correcting for the width-luminosity relation, and
taking the uncorrected zero point, ${\cal M}_B = -3.17 \pm 0.03$ of
Hamuy {\em et~al.} (1996).   We find essentially the same results: 
$\Omega_{\rm M} = 0.93 \;^{+0.69}_{-0.60}$ for $\Lambda = 0$ and
$\Omega_\Lambda = 0.03 \;^{+0.29}_{-0.34}$ for a flat $\Omega_{\rm
total}  = 1$ universe. The measurement uncertainty for the analysis
using light-curve-width  correction is not significantly better than 
this one, without correction, because for this particular data set the
smaller  calibrated dispersion, $\sigma_{M_B,{\rm corr}}$ as opposed to
$\sigma_{M_B}$, is offset by the uncertainties in the measurements of
the light curve widths. The goodness of fit, however, for  the
uncorrected  magnitudes is lower,  $Q(\chi^2|\nu)= 0.34$, for both the
$\Lambda=0$ and flat universe.  This indicates that the width-luminosity
relation provides a  better model, although this was already seen in the
improved RMS dispersion of the corrected magnitudes. Table~\ref{tabletwo}
summarizes the results.

\placetable{tabletwo}

As a cross-check, we also analyzed the results for all seven supernovae,
not ``width-corrected,'' including the two that had measured widths
outside the range of the low-redshift width correction.  We find 
$\Omega_{\rm M} = 0.70 \;^{+0.57}_{-0.49}$ for $\Lambda = 0$ and
$\Omega_\Lambda = 0.15 \;^{+0.24}_{-0.28}$ for  $\Omega_{\rm total}  =
1$.   
 
\begin{figure}[tbp]
\psfig{figure=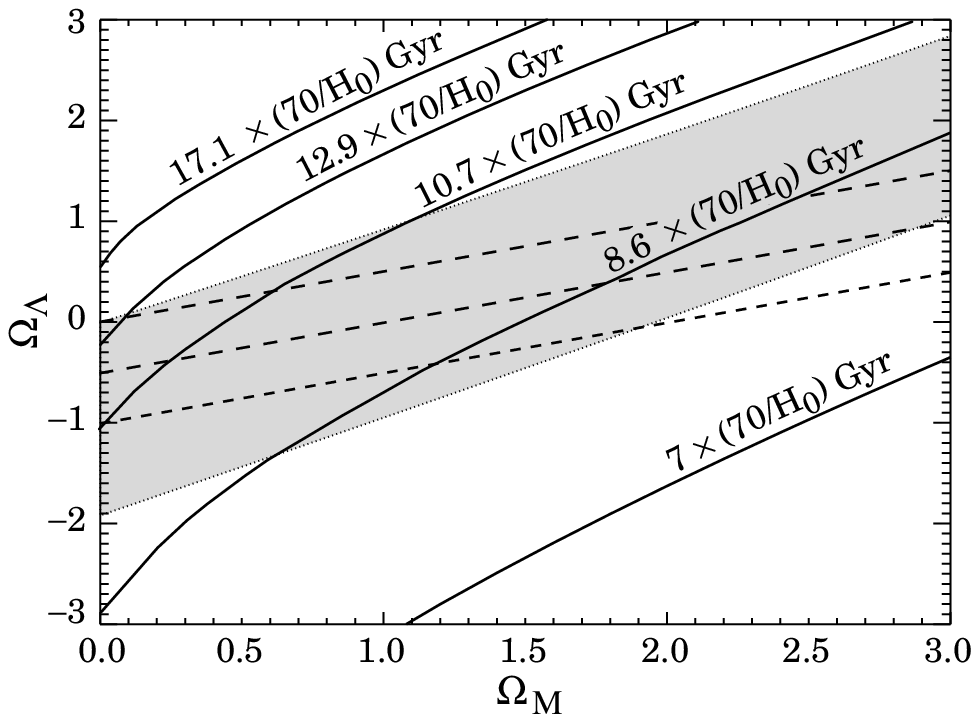,height=4.2in}
\caption[lvsm]{
Contour plot of the 1$\sigma$ (68\%) confidence
region  in the $\Omega_{\Lambda}$ versus   $\Omega_{\rm M}$ plane, for
the first seven high-redshift supernovae. The solid lines show contours
for the age of the universe (in Gyr), normalized to $H_0 = 70$ in units
of km s$^{-1}$ Mpc$^{-1}$. The dashed lines are the contours of constant
deceleration parameter $q_0$ = 0.0 (top), 0.5 (middle), 1.0 (bottom).}
\label{agelimit}
\end{figure}

We emphasized in Section 2 that the  1$\sigma$  confidence region of 
Figure~\ref{omegaomega} is not parallel to the contours of constant
$q_0$.  For comparison, the $q_0$ contours are drawn (dashed lines) in
Figure~\ref{agelimit}.  The closest $q_0$ contour  varies from $\sim$0.5
to $\sim$1 in the region with $\Omega_{\rm M} \le 1$. Any single value
of $q_0$ would thus be only a rough approximation to the true confidence
interval dependent on $\Omega_{\rm M}$ and $\Omega_\Lambda$  shown in
Figure~\ref{agelimit}.

\section{Checks for Sources of Systematic Error}

Since this is a first measurement of cosmological parameters using SNe
Ia, we list here some of the more important concerns, together with the
checks and tests that would address them.   By considering each of the
separate subsamples of SNe that avoid a particular source of bias, we
can show that no one source of systematic error alone can be responsible
for the measured values of $\Omega_{\rm M}$ and $\Omega_\Lambda$.    We
cannot, however, exclude a conspiracy of biases each moving an
individual supernova's  measured values into agreement with this result. 
Fortunately, this sample of seven  supernovae is only the beginning of a
much larger data set of SNe Ia at high redshifts, with more complete
multiband light curves and spectral coverage.

{\em Width-Brightness Correction.} 
For the present data set, the measurements  of $\Omega_{\rm M}$ and
$\Omega_\Lambda$ are the same whether or not we apply the light-curve
width-luminosity correction, and so do not depend upon our confidence in
this empirical calibration.  This is due to the similar distribution of 
light curve widths for the high-redshift supernovae and the low-redshift
supernovae used for calibration.  

For future data sets, it is possible that the width distribution will
differ, for example if  we were to find more supernovae in clusters of
ellipticals and confirm the tendency to  find narrower/fainter
supernovae in ellipticals.  (Note that SN~1994am,  in an elliptical
host,  does in fact have a somewhat narrow light curve, with $s=0.84$.)
For such a data set the result would depend on the validity of the
width-luminosity correction.   Such a correction dependence could be
checked by restricting the analysis to the supernovae that pass the
Vaughan {\em et~al.} (1995) $B$$-$$V$ color test for ``normals,''  and
then applying no width-luminosity correction. 

{\em Extinction.}
While the extinction due to our own Galaxy has been incorporated in this
analysis, we do not know the SN host-galaxy extinctions.  Note that 
correcting for any neglected extinction for the high-redshift supernovae 
would tend to brighten our estimated SN effective magnitudes and hence
move the best fit of Equation~\ref{simplemagz} towards even higher
$\Omega_{\rm M}$ and lower $\Omega_\Lambda$ than the current results. 
We can check that the extinction is not strongly affecting the results
by considering two supernovae for which there is evidence against
significant extinction. SN~1994am  is in an elliptical galaxy, and for
SN~1994G   there is the previously-mentioned $R$$-$$I$ color that
provides evidence against significant reddening.   The best fit  values
for this two-SN subset are consistent with that of the full set of  SNe:
$\Omega_{\rm M} = 1.31 \;^{+1.23}_{-0.98}$ for $\Lambda = 0$ and
$\Omega_\Lambda = -0.15 \;^{+0.47}_{-0.61}$ for  $\Omega_{\rm total}  =
1$.

If there were uncorrected host-galaxy extinction for the  {\em
low-redshift} supernovae used to find the magnitude zero point ${\cal
M}$, this would lead to an opposite bias. The 18 Cal\'{a}n/Tololo
supernovae, however, all have unreddened colors ($B-V < 0.2$~mag).  For
the range of widths of these supernovae  ($0.8 \ls \Delta m_{15} \ls
1.75$~mag), the range of intrinsic color at maximum light should be 
$-0.05 \ls B-V \ls 0.5$~mag, so the color excess is  limited to $E(B-V)
< 0.25$~mag. A stronger constraint can be stated for the seven of these
18 Cal\'{a}n/Tololo supernovae that were included in a sample studied by
Riess {\em et al.} (1996), who used the multicolor correction-template
method to estimate extinction.  They found that only one of these seven
supernovae showed any significant extinction.  (For that one supernova,
SN~1992P, they found $A_V = 0.11$~mag.)   It is thus unlikely that
host-galaxy extinction of the Cal\'{a}n/Tololo supernovae is strongly
biasing our results, although it will be useful to test the remaining
supernovae of this set  for evidence of extinction, using the multicolor
correction-template method.  (It should be noted that the method of
Riess {\em et al.} estimates total extinction values for several
supernovae that are significantly less than the Galactic extinction
estimated by Burstein \& Heiles 1982, hence some caution is still necessary
in interpreting these results.) 

A general argument can be made even without these color measurements:  
the intrinsic magnitude dispersions, $\sigma_{M_B,{\rm corr}}^{\rm
Hamuy} = 0.17$ or our value $\sigma_{M_B,{\rm corr}}=0.19$~mag,  provide
an upper bound on the typical extinction present in  either the low
redshift or the high redshift supernova samples, since a broad range of
host-galaxy extinction (which would have a larger mean extinction) would
inflate these dispersions.  We use Monte Carlo studies to bound the
amount of  bias due to a distribution of extinction that would inflate 
a hypothetical intrinsic dispersion of $\sigma_{M_B}^{\rm hypothesis} =
0.13$  to  the values actually measured, $\sigma_{M_B,{\rm corr}}^{\rm
Hamuy} = 0.17\pm0.04$ and   $\sigma_{M_B,{\rm corr}}=0.19\pm0.08$~mag. 
We find, at the 90\% confidence level, less than 0.06 magnitude  bias
towards higher $\Omega_{\rm M}$ and lower $\Omega_\Lambda$ and less than
0.10 magnitude bias towards lower $\Omega_{\rm M}$ and higher
$\Omega_\Lambda$. These measurements and bounds on extinction will be
even more important as we study supernovae at still higher redshifts and
need to check for evolution in the host galaxy dust.

We choose high Galactic latitude fields whenever possible, so that the
extinction correction for our own Galaxy and its uncertainty are
negligible.  The one major exception in this current data set is SN
1994al,  with $A_R \approx 0.23$~mag. (This supernova also appears to be
a somewhat fainter outlier among the  width-corrected magnitudes on
Figure~\ref{magredshift}b.) Since there is more uncertainty in the
Galactic extinction correction for this supernova, we have fit just the
other four width-corrected supernova magnitudes. For SN~1994H, SN~1994am, 
SN~1994G,  and SN~1994an, we find $\Omega_{\rm M} = 1.25
\;^{+0.82}_{-0.69}$ for $\Lambda = 0$ and $\Omega_\Lambda = -0.12
\;^{+0.33}_{-0.40}$ for  $\Omega_{\rm total}  = 1$.

{\em K Correction.}
The generalized $K$ correction used to transform $R$-band magnitudes of
high redshift SNe to $B$-band rest frame magnitudes was tested for a
variety of SN~Ia spectra and found to vary by less than 0.04 mag for
redshifts $z < 0.6$ (Kim, Goobar, \& Perlmutter 1996).  The test spectra
sample included examples in the range of light-curve widths between
$\Delta m_{15} = 0.94$  mag (for SN~1991T)  and $\Delta m_{15} = 1.47$
mag (for  SN~1992A), but $K$ corrections still need to be calculated to
check for possible difference outside of this regime of the SN~Ia
family.   Two of the five supernovae we used in our measurement fall
just outside of this range of widths.  As a check of a possible
systematic error due to this, we consider just the subsample of three
supernovae, SN~1994al, SN~1994am, and SN~1994G, with light curve widths
within the studied range of $K$ corrections. We find  $\Omega_{\rm M} =
0.61 \;^{+0.89}_{-0.61}$ for $\Lambda = 0$ and $\Omega_\Lambda = 0.18
\;^{+0.35}_{-0.43}$ for  $\Omega_{\rm total}  = 1$.

{\em Malmquist Bias.}
The tendency to find the most luminous members of a distribution at
large distances in a magnitude-limited search would appear  to bias our
results towards larger values of  $\Omega_{\rm M}$ and lower values of
$\Omega_\Lambda$. However, as the SN~Ia Hubble diagram uses the peak
brightness rather than that at detection, and the supernovae are
``corrected'' from different intrinsic brightnesses, any such Malmquist
bias would not operate the same way it does  for a population of
normally-distributed static standard candles.   The key issue is whether
supernova samples are strongly biased at the detection level, as would
be the case if  most supernovae were discovered close to the threshold
value. Our detection-efficiency studies (see Pain {\em et~al.} 1996)
show that the seven high-redshift supernovae were detected approximately
0.5 to 2 magnitudes  brighter than our limiting (50\% efficiency)
detection magnitude, $m_{_{d{-}50}}$ (see the final row of
Table~\ref{tableone}), 
and the efficiency on our CCD images  drops off slowly beyond this
limit.  (Several of the SNe that we find at a significantly brighter 
magnitude than our threshold are in clusters that are closer than the
limiting distance for a SN~Ia.  Such  inhomogeneities in galaxy
distribution can lead to a sample of supernovae that are not distributed
primarily near the magnitude limit as  expected in a magnitude limited
search.) In contrast, the Cal\'{a}n/Tololo survey detected most of the
low-redshift supernovae within $\sim$0.7 magnitude of detection
threshold and their efficiency on photographic plates dropped quickly
beyond that limit (Maza, private communication).  Malmquist bias may
therefore result, counter-intuitively,  in a slightly more luminous
sample of the intrinsic SN~Ia distribution for the low-redshift
photographic search than for the high-redshift CCD search.

Since, however,  our current results already suggest a relatively high
value for  $\Omega_{\rm M}$ and low value for  $\Omega_\Lambda$, we have
nonetheless checked the possibility of Malmquist bias distorting the
distributions of magnitudes we find at  high redshift.  First, as a
rough test, we have compared the results (uncorrected for light-curve
width) for the three supernovae discovered closest to the
50\%-efficiency detection threshold, $m_{_{d-50}}$,  to the three
supernovae discovered farthest from the threshold.   We find that the
difference in measured $\Omega_{\rm M}$ and $\Omega_\Lambda$ between
these two subsets is not significant, and it is opposite to the 
direction of Malmquist bias.

A more detailed quantitative study was based on a Monte Carlo analysis,
in which the detection efficiency curve for each supernova was used with
the Cal\'{a}n/Tololo ``corrected'' magnitude dispersion,
$\sigma_{{M}_B,{\rm corr}}^{\rm Hamuy}$, to estimate the magnitude bias
that should be present for each of the seven fields on which we
discovered high-redshift supernovae.  We find that only one supernova,
SN~1994G,  is on a field that shows any significant magnitude bias. 
Even for this supernova the bias, 0.01 mag, is still well below the
intrinsic dispersion of the supernovae.  Hence, when we reanalyze the
data set, correcting SN~1994G by this amount, we find an insignificant
change in our results for $\Omega_{\rm M}$ and $\Omega_\Lambda$.

Because the high-redshift supernovae include both intrinsically narrow,
subluminous cases and intrinsically broad, overluminous cases at
comparable redshifts, a simple cross-check  for Malmquist bias that is
independent of detection-efficiency studies can  be made by comparing
the results for these two subsets separately.  Malmquist bias would
affect these two subsets differently, leading to a higher $\Omega_{\rm
M}$ and lower  $\Omega_\Lambda$ for the broad, overluminous subsample 
than for the narrow, subluminous subsample.  With our current sample we
can compare only  two supernovae in each of these subsamples that are in
the ``correctable'' range of  $\Delta m_{15}$, so this will be a
particularly interesting test to   apply to the full sample of 
high-redshift supernovae when their light-curve observations are
completed.   The current data show no evidence of Malmquist bias.

{\em Supernova Evolution.}
Although there are theoretical reasons to believe that the physics of
the supernova explosion should not depend strongly on the  evolution of
the progenitor and its environment,  the empirical data are the final
arbiters.   Both the low-redshift and high-redshift supernovae are
discovered in a variety of host galaxy types, with a range of histories. 
The small dispersion in intrinsic magnitude across this range,
particularly after the width-luminosity correction, is itself an
indication that any evolution is not changing the relationship between
the light curve width/shape and its absolute brightness.  (Note that the
one supernova in an identified elliptical galaxy gives results for the
cosmological parameters consistent  with the full sample of supernovae;
such a comparative test will of course be more useful with the larger
samples.) In the near future, we will be able to look directly for signs
of  evolution in the $>$18 spectra observed for the larger sample of
high-redshift supernovae. So far, the spectral features studied match
the  low-redshift supernova spectra for the appropriate  day on the
light curve (in the supernova rest frame), showing no evidence for
evolution.  A more detailed analysis will soon be possible, as the host
galaxy spectra are observed after the supernovae fade, and it becomes
possible to study the supernova spectra without galaxy contamination.

{\em Gravitational Lensing.}
Since the mass of the universe is not homogeneously distributed,
there is a potential source of  increased magnitude
dispersion, or even a magnitude shift, due to
overdensities (or  underdensities) acting as gravitational
lenses that amplify (or deamplify) the supernova light.  This effect
was analyzed  in a simplified ``swiss cheese'' model by
Kantowski,  Vaughan, \& Branch (1995), and more recently
using a perturbed Friedmann-Lema\^{\i}tre cosmology  by Frieman (1996)
and an $n$-body simulation of a $\Lambda$CDM cosmology by
Wambsganss {\em et~al.} (1996).  The conclusion of the
most recent analyses is that the additional dispersion
is negligible at the redshifts
$z < 0.5$ considered in this paper:  Frieman estimates an upper limit
of less than 0.04 magnitudes in additional dispersion.  Any systematic
shift in magnitude distribution is similarly small: Wambsganss {\em et~al.}
give the difference between the median of the
distribution and the true value for their particular
mass-density distribution, which corresponds to a $\sim$0.025 magnitude
shift at $z=0.5$.  We can take this as a bound on the magnitude
shift, since our {\em averaged} results should be closer to the true
value than the median would be.  Alternative models for the
mass density distribution must satisfy the same observational
constraints on dark matter power at small scales (from pairwise
peculiar velocities and abundances of galaxy clusters) and therefore
should give similar results.

\section{Discussion}

We wish to stress two important aspects of this measurement. First,
although we have considered many potential sources of error and possible
approaches to the analysis, the essential results that we find are
independent of almost all of these complications:  this direct
measurement of the cosmological parameters can be derived from just the
peak magnitudes of the supernovae, their redshifts, and the
corresponding $K$ corrections.  

Second, we emphasize that the high-redshift supernova data sets provide
enough detailed information for each individual supernova that we can
perform tests for many possible sources of systematic error by comparing
results for supernova subsets that are affected differently by the
potential source of error.  This is a major benefit of these distance
indicators.

There have been many previous measurements   and limits for the mean
mass density of the Universe.   The methods that sample the largest
scales via peculiar velocities of galaxies and their production through
potential fluctuations have tended to yield values of $\Omega_{\rm M}$
close to unity (Dekel, Burstein, \& White 1996 but see Davis {\em et~al.} 1996 and Willick
{\em et~al.} 1997).  Such techniques, however, only constrain
$\beta=\Omega_{\rm M}^{0.6}/b$  where $b$ is the biasing factor for the
galaxy tracers, which is generally unknown (but see Zaroubi {\em et~al.}
1996).  Alternative methods based on cluster dynamics (Carlberg {\em
et~al.} 1995) and gravitational lensing (Squires {\em et~al.} 1996) give
consistently lower values of $\Omega_{\rm M} \simeq$0.2-0.4. There are
many reasons, however, why these values could be underestimates,
particularly if clusters are surrounded by extensive dark halos. The
measurement using supernovae has the advantage of being a global
measurement that includes all forms of matter, dark or visible, baryonic
or non-baryonic, ``clumped'' or not.  Our  value of   $\Omega_{\rm M} =
0.88 \;^{+0.69}_{-0.60}$ for a $\Lambda = 0$ cosmology does not yet rule
out many currently favored theories, but it does provide a first global
measurement that is consistent with  a standard Friedmann cosmology. Our
measurement for a flat universe, $\Omega_{\rm M} = 0.94
\;^{+0.34}_{-0.28}$, is more constraining:  even at the 95\% confidence
limit, $\Omega_{\rm M} > 0.49$,  this result  requires non-baryonic dark
matter.

In a flat universe ($\Omega_{\rm total} =1 $), the constraint on the
cosmological constant from gravitational lens statistics is
$\Omega_\Lambda  < 0.66$ at the 95\% level (Kochanek 1995; see also Rix
1996 and Fukugita {\em et~al.} 1992).    (In establishing this limit,
Kochanek included the statistical  uncertainties in the observed
population of lenses, galaxies, quasars, and the uncertainties in the
parameters relating galaxy luminosity to dynamical variables;  he also
presented an analysis of the various models of the lens, lens galaxy,
and extinction to show the result to be insensitive to the  sources of
systematic error considered.) Our upper limit for the flat universe
case,  $\Omega_{\Lambda} < 0.51$ ( 95\% confidence level),  is somewhat
smaller than the gravitational lens upper limit,  and provides an
independent, direct route to this important result.  The supernova
magnitude-redshift approach actually provides a measurement,
$\Omega_\Lambda = 0.06 \;^{+0.28}_{-0.34}$, rather than a limit. The
probability of gravitational lensing varies steeply with
$\Omega_\Lambda$ down to the $\Omega_{\Lambda} \approx 0.6$ level, but
is less sensitive below this level, so the limits from gravitational
lens statistics are unlikely to improve dramatically.  The uncertainty
on  the $\Omega_\Lambda$ measurement from the high-redshift supernovae,
however, is likely to narrow by $\sqrt 3$ as we reduce the data from the
next $\sim$18  high-redshift supernovae that we are now observing.

For the more general case of a Friedmann-Lema\^{\i}tre cosmology with
the sum of $\Omega_{\rm M}$ and $\Omega_\Lambda$ unconstrained, the
measured bounds on the cosmological constant are, of course, broader. A
review of the observational constraints on the cosmological constant by
Carroll, Press, \& Turner (1992) concluded that the best bounds are $-7
< \Omega_\Lambda  < 2$, based on the existence of high-redshift objects,
the ages of globular clusters and cosmic nuclear chronometry, galaxy
counts as a function of redshift or apparent magnitude, dynamical tests
(clustering and structure formation), quasar absorption line statistics,
gravitational lens statistics, and the astrophysics of distant objects. 
Even in the case of unconstrained $\Omega_{\rm M}$ and $\Omega_\Lambda$,
we find a significantly tighter bound on  the lower limit:
$\Omega_\Lambda  > -2.3$ (at the two-tailed 95\% confidence level, for
comparison).   Our upper limit is similar to the previous bounds, but
provides an independent check: $\Omega_\Lambda  < 1.1$ for  $\Omega_{\rm
M}  \ls 1$ or $\Omega_\Lambda  <2.1$ for  $\Omega_{\rm M}  \ls 2$.

These current measurements of  $\Omega_{\rm M}$ and $\Omega_\Lambda$ are
inconsistent with high values of the Hubble constant ($H_0 > 70$ km
s$^{-1}$ Mpc$^{-1}$) taken together with  ages greater than $t_0=13$
billion years for the globular cluster stars. This can be seen in
Figure~\ref{agelimit}, which compares the  confidence region for the
high-redshift supernovae on the $\Omega_{\rm M}$--$\Omega_\Lambda$ 
plane with the $t_0$ contours.   Above the 68\% confidence region, only
a small fraction of probability lies at ages 13 Gyr or higher for $H_0
\ge 70$ km s$^{-1}$ Mpc$^{-1}$, and this age  range favors low values
for $\Omega_{\rm M}$.  Once again, the analysis of our next set of
high-redshift supernovae will test and refine these results.

\acknowledgements

We wish to acknowledge the community effort of many observers who
contributed to the data, including Susana Deustua, Bruce Grossan,  Nial
Tanvir, Emilio Harlaftis, J. Lucy, J. Steel, George Jacoby, Michael
Pierce, Chris O'Dea, Eric Smith, Wilfred Walsh,  Neil Trentham, Tim
Mckay, Marc Azzopardi, Jean-Paul Veran, Roc Cutri,  Luis Campusano,
Roger Clowes, Matthew Graham, Luigi Guzzo, and Pierre Martin. We thank
Marc Postman, Tod Lauer, William Oegerle, and John Hoessel, in
particular, for their more extended participation in the observing, in
coordination with the Deeprange Project. We thank David Branch and Peter Nugent for
discussions and comments.   The observations described in this paper
were primarily obtained as visiting/guest astronomers at the Isaac
Newton and William Herschel Telescopes, operated by the Royal Greenwich
Observatory at the Spanish Observatorio del Roque de los Muchachos of
the Instituto de Astrofisica de Canarias; the Kitt Peak National
Observatory 4-meter and 2.1-meter telescopes and Cerro Tololo
Inter-American Observatory 4-meter telescope, both operated by the
National Optical Astronomy Observatory under contract to the National
Science Foundation; the Keck I 10-m telescope, operated by the California
Association for Research in Astronomy
as a scientific partnership between
the California Institute of Technology and the University of
California and made possible by the generous gift of the W. M.
Keck Foundation;  the Nordic Optical 2.5-meter
telescope; and the Siding Springs 2.3-meter telescope of the Australian
National University.  We thank the staff of these observatories for
their excellent  assistance. This work was supported in part by the
Physics Division,  E.~O. Lawrence Berkeley National Laboratory of the
U.~S. Department of Energy under Contract No. DE-AC03-76SF000098, and by
the National Science Foundation's Center for Particle Astrophysics,
University of California, Berkeley under grant No. ADT-88909616.  
A.~V.~F. acknowledges the support of NSF grant No. AST-9417213 and A.~G.
acknowledges the support of the Swedish Natural Science Research
Council.


\clearpage

\clearpage

\begin{deluxetable}{lccccccc}
\footnotesize
\tablecaption{Supernova Data and Photometry Error Budget\label{tableone}}
\tablehead{
\colhead{} &
 \colhead{SN 1992bi}    &
 \colhead{SN 1994H}       &   \colhead{SN 1994al}  &
 \colhead{SN 1994F}       &   \colhead{SN 1994am} &
 \colhead{SN 1994G}       &   \colhead{SN 1994an}
 }
\startdata
{$z\;^{[a]}$} & 0.458 & 0.374 & 0.420 &  0.354 & 0.372 & 0.425 & 0.378 \cr
{$m_R$}   & 22.01 (9)$^b$ &  21.38 (5) & 22.42 (6) & 22.06 (17) & 21.73 (6)
&21.65 (16) & 22.02 (7) \cr
 {$A_R\;^{[c]}$}  &0.003 (1) & 0.039 (4) & 0.228 (114) & 0.010 (1) 
 & 0.039 (4) & 0.000 (1) & 0.132 (13) \cr
{$K_{BR}$}   &  $-$0.70 (1) & $-$0.58 (3) &  $-$0.65 (1) & $-$0.56 (3)  &  
 $-$0.58 (2.5) &  $-$0.66 (1) & $-$0.59 (2.5) \cr
{$m_B$} & 22.71 (9)&  21.92 (6)& 22.84 (13) & 22.61 (18) &  22.27 (7) 
& 22.31 (16) & 22.48 (7) \cr
\cr

{$s$}  & 1.45 (18) &   1.09 (5) &   0.95 (12) &  0.67 (15) & 
   0.86 (3) &    1.01 (13) &   0.77 (9) \cr
$\Delta m_{15}$ & 0.47 (17) &  0.91 (9)  &  1.17 (27)  &  2.04 (65) &  1.39 (10) &  1.04 (25) & 1.65 (32) \cr
{$\Delta_{\rm corr}^{\scriptscriptstyle\{1.1\}}$} &  [{\it 0.55 (20)}]$^d$ &  0.16 (9) &  $-$0.06 (23) &  [{\it $-$0.81 (59)}]$^d$ 
&  $-$0.25 (10) & 0.05 (22)
&  $-$0.47 (30) \cr 
{$m_{B,{\rm corr}}$} & [{\it 23.26 (24)}]$^d$ &  22.08 (11)   &  22.79 (27) &  [{\it 21.80 (69)}]$^d$ 
&  22.02 (14) &  22.36 (35)  &  22.01 (33)  \cr
\cr

$t_{\rm spect}\,^{[e]}$  & 84 (3)  &  219 (2)  &  209 (2)$^f$  & -2 (2)$^g$ &203 (2)$^f$& 13 (1)$^h$ 
& 3 (2) \cr
{Bands}  & $R$ & $B,R$ &  $R$ &  $R$ &  $R$ & $B,R,I$ 
& $B,R$ \cr
{$m_{_{d-50}}\;^{[j]}$} & 1.0 & 1.9 & 0.8 & 0.4 & 0.9 & 0.5 & 1.5 \cr
\enddata

\tablecomments{The uncertainties in the least significant digit are
given in parentheses.  Section 4 of the text defines the variables.}
\tablenotetext{a}{The redshifts were measured from host-galaxy spectral
features, and their uncertainties are all $\ls$0.001.}

\tablenotetext{b}{This value for $m_R$ of SN~1992bi incorporates more
recent light curve and calibration data than were available when
Perlmutter {\em et~al.} (1995) was prepared.   These new data have
improved the photometry error bars, as that paper suggested.  The small
change in the $m_R$ value quoted is also partly due to the extra degree
of freedom in the new fit of the template light curve to a stretch
factor, $s$.}

\tablenotetext{c}{Extinction for our Galaxy, $A_R = {\cal R}_R E(B$$-
$$V)_{\rm B\&H}$, where  ${\cal R}_R = 2.58 \pm 0.01$ was calculated
using well-observed SN~Ia spectra redshifted appropriately for each
supernova,  and the $E(B$$- $$V)_{\rm B\&H}$ values for each supernova
coordinate are from Burstein \& Heiles (1982) and Burstein, private
communication (1996).}

\tablenotetext{d}{Since SN~1992bi and SN~1994F have $\Delta m_{15}$
values outside of the range for low-redshift SNe, we do not use these
lightcurve-width-corrected values  for $m_{B,{\rm corr}}$ for any of the
results of this paper. In the figures,  we do plot them as grey squares,
along with white squares for  the following  alternative values that are
corrected only to the extreme of $\Delta_{\rm
corr}^{\scriptscriptstyle\{1.1\}}$ values for low-redshift supernovae:
$m_{B,{\rm corr}}^{\rm extreme} = 22.97 (24)$ for SN~1992bi,  and
$m_{B,{\rm corr}}^{\rm extreme} = 22.05 (69)$ for SN~1994F.}
\tablenotetext{e}{$t_{\rm spect}$ is the number of days past $B$
maximum in the  supernova rest frame of the first spectrum.  Generally,
only the host galaxy spectrum is bright enough to be useful later than
$t_{\rm spect} \gs 30$ days.}

\tablenotetext{f}{Spectrum of host galaxy showed the strong 4000\AA\
break of an elliptical galaxy, indicating that SN~1994al and SN~1994am
are type Ia supernovae (see also Figure~\ref{morphology}).}

\tablenotetext{g}{Keck spectrum of SN~1994F observed by J.B. Oke, J.
Cohen, and T. Bida.}

\tablenotetext{h}{MMT spectrum observed by P. Challis, A. Riess, and
R. Kirshner. We also observed a spectrum at the Keck telescope 15
rest-frame days past $B$ maximum.}

\tablenotetext{j}{$m_{_{d-50}}$ is the magnitude difference between the
discovery magnitude  and the  detection threshold, i.e. the magnitude at
which 50\% of simulated supernovae are detected for the image in which
the supernova was discovered.  Note that the discovery magnitude is
generally fainter than the peak magnitude, $m_R$, since most discoveries
are before maximum light.}

\end{deluxetable}

\begin{deluxetable}{lcc}
\footnotesize
\tablecaption{Cosmological Parameters 
from SNe 94H, 94al, 94am, 94G, and 94an\label{tabletwo}}
\tablewidth0pt
\tablehead{\colhead{}
      &\colhead{$\Omega_{\rm M} $} &  \colhead{$\Omega_\Lambda$} \cr
& \colhead{for  no-$\Lambda$ universe}&  \colhead{for flat universe}\cr
 &  \colhead{($\Omega_\Lambda =0$)}&    
\colhead{($\Omega_{\rm total} \equiv \Omega_{\rm M} +\Omega_\Lambda = 1$)}}
\startdata
Light-Curve-Width Corrected$^a$ & $0.88 \;^{+0.69}_{-0.60}$ 
& $0.06 \;^{+0.28}_{-0.34}$ \cr
Uncorrected$^b$ & $0.93 \;^{+0.69}_{-0.60}$ 
& $0.03 \;^{+0.29}_{-0.34}$\cr
\enddata

\tablenotetext{a}{The results are the best fit of the
lightcurve-width-corrected data, $m_{B,{\rm corr}}  = m_B + \Delta_{\rm
corr}^{\scriptscriptstyle\{1.1\}}$, of Table~\ref{tableone} to
Equation~\ref{simplemagz}, with ${\cal M} = {\cal M}_{B,{\rm
corr}}^{\scriptscriptstyle\{1.1\}} = -3.32 \pm 0.05$. The probability of
obtaining the best-fit $\chi^2$ or higher for both the $\Lambda=0$ and
flat universes is  $Q(\chi^2|\nu)= 0.75$.}

\tablenotetext{b}{The results are the best fit of the data, $m_B$, of
Table~\ref{tableone} to Equation~\ref{simplemagz}, with ${\cal M} =
{\cal M}_B = -3.17 \pm 0.03$. The probability of  obtaining the best-fit
$\chi^2$ or higher for both the $\Lambda=0$ and flat universes is 
$Q(\chi^2|\nu)= 0.34$.}

\end{deluxetable}

\end{document}